# EFFICIENT INDEPENDENT COMPONENT ANALYSIS

By Aiyou Chen[1] and Peter J. Bickel[2]

*Bell Labs and University of California, Berkeley*

Independent component analysis (ICA) has been widely used for blind source separation in many fields such as brain imaging analysis, signal processing and telecommunication. Many statistical techniques based on M-estimates have been proposed for estimating the mixing matrix. Recently, several nonparametric methods have been developed, but in-depth analysis of asymptotic efficiency has not been available. We analyze ICA using semiparametric theories and propose a straightforward estimate based on the efficient score function by using B-spline approximations. The estimate is asymptotically efficient under moderate conditions and exhibits better performance than standard ICA methods in a variety of simulations.

**1. Introduction.** Independent component analysis (ICA) aims to separate independent blind sources from their observed linear mixtures without any prior knowledge. This technique has been widely used in the past decade to extract useful features from observed data in many fields such as brain imaging analysis, signal processing and telecommunication. Hyvarinen, Karhunen and Oja [16] described a variety of applications of ICA. For example, Vigario, Jousmaki, Hamalainen, Hari and Oja [25] used ICA to separate artifacts from magnetoencephalography (MEG) data, without the burden of modeling the process that generated the artifacts.

Standard ICA represents an $m \times 1$ random vector $X$ (e.g., an instantaneous MEG image) as linear mixtures of $m$ mutually independent random variables $(S_1, \ldots, S_m)$ (e.g., artifacts and other brain activities), where the distribution of each $S_i$ is totally unknown. That is, for $S = (S_1, \ldots, S_m)^T$ and some $m \times m$ nonsingular matrix $W$,

$$(1.1) \qquad X = W^{-1}S.$$

Received July 2003; revised February 2006.

[1]Supported in part by NSF Grant DMS-01-04075 and Lucent Technologies.

[2]Supported in part by NSF Grant DMS-01-04075.

*AMS 2000 subject classifications.* Primary 62G05; secondary 62H12.

*Key words and phrases.* Independent component analysis, semiparametric models, efficient score function, asymptotically efficient, generalized M-estimator, B-splines.

Here, $W^{-1}$ is called the *mixing matrix*. Given $n$ i.i.d. observations, $X^1, \ldots, X^n$, from the distribution of $X$, the aim is to estimate $W$ and, thus, to separate each component of $S = WX$ such that the components are maximally mutually independent. $W$ is called the *unmixing matrix*. This can be seen as a projection pursuit problem [14] in which $m$ directions are sought such that the corresponding projections are most mutually independent.

It was shown by Comon [9] that $W$ is identifiable up to scaling and permutation of its rows if at most one $S_i$ is Gaussian [18]. Model (1.1) can be viewed as a semiparametric model with parameters $(W, r_1, \ldots, r_m)$, where $r_i$ is the probability density function (PDF) of $S_i$. Our interest centers on $W$; $(r_1, \ldots, r_m)$ are nuisance parameters.

ICA was motivated by neurophysiological problems in the early 1980's (see [16]) and two classes of methods have been proposed to estimate $W$. One class involves specifying a particular parametric model for each $r_i$ and then optimizing contrast functions that involve $(W, r_1, \ldots, r_m)$. Primary examples of this approach are maximum likelihood (ML) (e.g., [19, 21]) or, equivalently, minimum mutual information (e.g., [9]), minimizing high-order correlation between components of $WX$ (e.g., [7]) and maximizing the non-Gaussianity of $WX$ (e.g., [15]). A second class of methods view ICA as a semiparametric model and assume nothing about the distributions of the components $S_i$. Thus, two distinct goals can be formulated: (i) to find estimates $\hat{W}$ of $W$ that are consistent or, even better, $\sqrt{n}$-consistent—that is, $\hat{W} = W + O_p(n^{-1/2})$ and (ii) to find procedures that achieve the information bound—that is, estimates of $W$ which are asymptotically normal and have smallest variance-covariance matrix among all estimates that are uniformly asymptotically normal in a suitable sense; see [5]. Amari [1] formally demonstrated that to achieve the information bound in this situation, a method must estimate the densities of the sources. In fact, it can even be shown [6] that for any fixed estimating equation corresponding to maximizing an objective function, there is a possible distribution of sources for which the global maximizer is inconsistent, despite the consistency of a local solution near the truth.

Recently, some nonparametric methods to estimate $W$ have appeared. For example, Bach and Jordan [3] proposed: (i) To reduce the dimension of the data by using a kernel representation and (ii) to choose $W$ so as to minimize the empirical *generalized variance* among the components of $WX$. Hastie and Tibshirani [13] proposed maximizing the penalized likelihood as a function of $(W, r_1, \ldots, r_m)$ and Vlassis and Motomura [26] proposed maximizing the likelihood by using Gaussian kernel density estimation. Various performance analyses have been carried out using simulations. The Vlassis–Motomura and Hastie–Tibshirani methods are of the same type as ours, but these papers do not provide a method for tuning the procedures and nothing has been proven about their asymptotic properties. Samarov and Tsybakov [22] proposed and analyzed a $\sqrt{n}$-consistent estimate of $W$ under mild conditions. Chen and Bickel [8] analyzed the method of Eriksson and Koivunen [12] based on characteristic functions and showed it to be consistent under the minimal identifiability conditions and $\sqrt{n}$-consistent under additional mild conditions. This paper concerns the construction of efficient estimates. We develop an efficient estimator by using efficient score functions after starting the algorithm at a consistent point based on the PCFICA algorithm of Chen and Bickel [8].

The outline of the paper is as follows. In Section 2 we analyze ICA as a semiparametric model and propose a new method to estimate $W$ using the efficient score function. The main theorem is given in Section 3. Numerical studies are given in Section 4. Technical details are provided in Sections 5 and 6.

*Notation.* In this paper, $W$ denotes an $m \times m$ matrix and $W_i$ and $W_{ij}$ denote the $i$th row and the $(i,j)$th element of $W$, respectively. $A^T$ denotes the transpose of a matrix $A$ and $A^{-T}$ denotes the transpose of $A^{-1}$. For any matrix $A$ with column vectors $\{\mathbf{a}_i : 1 \leq i \leq k\}$, $\|A\|_F = \sqrt{tr(A^T A)}$ and $vec(A) = (\mathbf{a}_1^T, \mathbf{a}_2^T, \ldots, \mathbf{a}_k^T)^T$, a column vector created from $A$. Define the *sup-norm* as $|f|_\infty = \sup_{t \in \mathbb{R}} |f(t)|$. $X^i$ denotes the $i$th random sample from the distribution of $X$. The population (empirical) law of $X$ is denoted by $P$ ($P_n$). $X_i$ and $S_i$ denote the $i$th element of $X$ and $S$, respectively. Denote the vector of density functions $(r_1, \ldots, r_m)$ by $r_{1:m}$. A vector or matrix of functions is denoted in boldface. For a vector of functions $\mathbf{B}$, $\mathbf{BB}^T(\mathbf{x})$ shall be used as an abbreviation of $\mathbf{B}(\mathbf{x})[\mathbf{B}(\mathbf{x})]^T$.

## 2. Semiparametric inference.

In this section, we first briefly review the salient features of estimation in semiparametric models and then show how to solve an approximate efficient score equation for estimating $W$ in the ICA model.

2.1. *Efficient estimates for semiparametric models.* Given a semiparametric model, $X^1, \ldots, X^n$ i.i.d. $\{P_{(\theta, \eta)} : \theta \in \Omega \subset \mathbb{R}^d, \eta \in \mathcal{E}\}$, where $\mathcal{E}$ is a subset of a function space, estimates $\hat{\theta}_n$ of $\theta$ are called *regular* if $\sqrt{n}(\hat{\theta}_n - \theta)$ converges in law uniformly in $P_{(\theta_n, \eta_n)}$, where $(\theta_n, \eta_n)$ converges to $(\theta_0, \eta_0)$ in a smooth way. Then if there is a regular estimate that is uniformly best (call it $\theta_n^*$), it must have the form

$$(2.1) \qquad \theta_n^* = \theta + \frac{1}{n} \sum_{i=1}^{n} \tilde{\mathbf{l}}(X^i, \theta, \eta) + o_p(n^{-1/2})$$

under $P_{(\theta, \eta)}$. The function $\tilde{\mathbf{l}}$ is called the *efficient influence function* in [5]. When $\eta = (\eta_1, \ldots, \eta_{d'})$ is a Euclidean parameter, $\tilde{\mathbf{l}}$ is, under regularity conditions, the influence function of the ML estimator (MLE) of $\theta$. That is, if $\dot{\mathbf{l}} = (\frac{\partial l}{\partial \theta^T}, \frac{\partial l}{\partial \eta_1}, \dots, \frac{\partial l}{\partial \eta_{d'}})^T$, where $l$ is the log-likelihood function of a single observation and $\mathbf{I}(\theta, \eta) \equiv E\dot{\mathbf{l}}\dot{\mathbf{l}}^T(X, \theta, \eta)$ is the Fisher information matrix, then $\tilde{\mathbf{l}}$ is the first $d$ coordinates of the vector $\mathbf{I}^{-1}(\theta, \eta)\dot{\mathbf{l}}$. An alternative formulation is to begin by defining the efficient score function $\mathbf{l}^* = (\mathbf{l}_1^*, \dots, \mathbf{l}_d^*)^T$ with

$$\mathbf{l}_k^* = \frac{\partial l}{\partial \theta_k} - \sum_{j=1}^{d'} a_{jk}(\theta, \eta) \frac{\partial l}{\partial \eta_j},$$

where $a_{jk}(\theta, \eta)$ minimizes $E(\frac{\partial l}{\partial \theta_k}(X, \theta, \eta) - \sum_{j=1}^{d'} a_{jk}(\theta, \eta) \frac{\partial l}{\partial \eta_j}(X, \theta, \eta))^2$. That is, $\mathbf{l}^*$ is the projection of $\frac{\partial l}{\partial \theta}(X, \theta, \eta)$ onto the orthocomplement of span$\{\frac{\partial l}{\partial \eta_j}(X, \theta, \eta) : 1 \le j \le d'\}$. Then

$$\tilde{\mathbf{l}} = (E[\mathbf{l}^* \mathbf{l}^{*T}(X, \theta, \eta)])^{-1} \mathbf{l}^*.$$

When $\eta$ is infinite-dimensional, the generalization of span$\{\frac{\partial l}{\partial \eta_j}(X, \theta, \eta) : 1 \le j \le d'\}$ is the tangent space. That is defined to be the closed linear span of $\{\frac{\partial l}{\partial \lambda}(X, \theta, \eta(\lambda))|_{\lambda=0} : \eta(0) = \eta$ and $\lambda \to \eta(\lambda)$ defines a smooth one-dimensional submodel $\{P_{(\theta, \eta(\lambda))} : |\lambda| < 1\}\}$ in $\mathcal{L}^2(P_{(\theta, \eta)})$. Now, $\mathbf{l}^*$ is again obtained by projection onto the orthocomplement of this span. An extensive discussion of tangent spaces and the geometric interpretation of formulas such as the one above is given in [5], Chapters 2 and 3. For many canonical semiparametric models including ICA, $\mathbf{l}^*$ can be computed; we sketch the argument in the Appendix. Suppose that for each $\theta$, an estimate $\hat{\eta}(\theta)$ is available and is at least consistent. Then the usual Taylor expansions suggest that the solution of the generalized estimating equation

$$(2.2) \qquad \sum_{i=1}^{n} \mathbf{l}^*(X^i, \theta, \hat{\eta}(\theta)) = 0$$

will have an influence function $\tilde{\mathbf{l}}$ and, hence, be efficient. These heuristics and others are discussed in Chapter 7 of [5]. Of course, more than consistency is needed and after calculating $\mathbf{l}^*$ in our case, validating that (2.2) leads to (2.1) for a suitable $\hat{\eta}(\theta)$ is the subject of Sections 3, 5, 6 and the Appendix. Note that if $\hat{\eta}(\theta)$ maximizes $\sum_{i=1}^{n} l(X^i, \theta, \eta)$, then (2.2) simply gives the profile maximum likelihood estimate discussed in [20]. In that case, (2.2) simplifies, becoming equivalent to

$$\sum_{i=1}^{n} \frac{\partial l}{\partial \theta}(X^i, \theta, \hat{\eta}(\theta)) = 0.$$

Unfortunately, such $\hat{\eta}(\theta)$ do not exist in the ICA model. Using $\mathbf{l}^*$ instead of $\frac{\partial l}{\partial \theta}$ in the estimating equation (2.2) permits a less demanding choice of $\hat{\eta}(\theta)$. These issues are discussed in detail in [5], Chapter 7. In this paper, we simply show that a $\hat{\theta}$ solving (2.2) for a particular $\hat{\eta}(\theta)$ does indeed satisfy (2.1) in a sufficiently uniform sense. Optimality of $\hat{\theta}$ then follows from the general theory given in Chapter 3 of [5].

This technique is different from the quasi-ML method which belongs to the first class of methods in the ICA literature reviewed in Section 1. This approach is to guess some shape $\eta_0$ for $\eta$ and then use ordinary ML. Of course, if $\eta_0$ is true, then the resulting estimate is asymptotically Gaussian and has smaller variance than the $\hat{\theta}$ we discuss. But, if $\eta_0$ is false, then the estimate can be inconsistent. The ICA algorithms used for comparison in Section 4 such as FastICA [Hyvarinen and Oja (1997)] and extended infomax [19] are of this type. Closest to ours in spirit among these is the method of Pham and Garrat [21]. They use parametric models such as logsplines (see Section 2.3) for the nuisance parameters. However, they propose solving the score equations rather than (2.2). More importantly, they do not suggest increasing the model dimension with $n$, do not give a method for selecting the number of knots of the splines and, hence, are subject to the inconsistency we have discussed.

The remainder of Section 2 shows how to implement the idea given in (2.2) for the ICA model. Technical analysis is carried out in Section 3.

2.2. *Further notation and assumptions.* Let $W_P$ be a nonsingular unmixing matrix such that $S = W_P X$ has $m$ mutually independent components. Without loss of generality, assume that $\det(W_P) > 0$. For any row vector $w \in \mathbb{R}^m$, let $f_w$ denote the PDF of $wX$ and $\phi_w$ denote the density score function defined by $\phi_w(t) = -\frac{\partial}{\partial t} \log f_w(t) I(f_w(t) > 0)$, where $I(.)$ is an indicator function.

In model (1.1), the order and scaling of rows of $W$ or components of $S$ must be constrained for $W$ to be identifiable. For scaling, we take each $S_i$ to have absolute median 1, that is, $P(|S_i| \leq 1) = \frac{1}{2}$ or, equivalently,

$$(2.3) \qquad 2 \int_{-1}^{1} r_i(s) \, ds = 1.$$

Even after this choice, the correct unmixing matrix requires $2^m m!$ choices due to sign changes and row permutations. This ambiguity can be resolved in various ways, but we avoid being specific by assuming that a consistent starting value is available for $W_P$, say PCFICA of Chen and Bickel [8]. Let $\kappa(s) = 2I(|s| \leq 1) - 1$. Then (2.3) is equivalent to

$$\int \kappa(S_i) \, dP = 0.$$

Equation (2.2), for our case, can be written

$$\sum_{i=1}^{n} \mathbf{l}^*(X^i, W, \hat{\Phi}_W) = 0,$$

TABLE 1
*Algorithm EFFICA*

| |
|---|
| 1.    Input data $\{X^1, \ldots, X^n\}$, an initial estimate $\hat{W}^{(0)}$ and B-spline basis functions $\mathbf{B}^{(k)} \equiv (B_1^{(k)}, \ldots, B_{n_k}^{(k)})^T$ for $k = 1, \ldots, m$; |

2.   For $j = 0, 1, \ldots$ until convergence:

   1) Set $\mathcal{S}_k^{(j)} \equiv \{\hat{W}_k^{(j)} X^i : i = 1, \ldots, n\}$ for $k = 1, \ldots, m$;

   2) Set for $k = 1, \ldots, m$,
$$\hat{\phi}_k^{(j)} = \gamma_k^{(j)T} \mathbf{B}^{(k)},$$
     where $\gamma_k^{(j)} = (\hat{E}_k^{(j)}[\mathbf{B}^{(k)} \mathbf{B}^{(k)T}])^{-1} \hat{E}_k^{(j)}[\mathbf{D}^{(k)}]$, $\mathbf{D}^{(k)}$ is the derivative function of $\mathbf{B}^{(k)}$ and $\hat{E}_k^{(j)}$ is empirical expectation w.r.t. $\mathcal{S}_k^{(j)}$;

   3) Set $\hat{\Phi}_W$ at $W = \hat{W}^{(j)}$:
$$\hat{\Phi}_W(\mathbf{s}) = (\hat{\phi}_1^{(j)}(s_1), \ldots, \hat{\phi}_m^{(j)}(s_m))^T$$

   4) Set for $k = 1, \ldots, m$,
$$\hat{u}_k^{(j)} = \hat{E}_k^{(j)}[\hat{\phi}_k^{(j)} \psi], \ \hat{v}_k^{(j)} = \hat{E}_k^{(j)}[\psi], \text{ where } \psi(s) = 2sI(|s| \leq 1),$$
$$(\hat{\sigma}_k^{(j)})^2 = \hat{E}_k^{(j)}[\varphi], \text{ where } \varphi(s) = s^2,$$
$$\hat{\alpha}_k^{(j)} = -\frac{(1 - \hat{u}_k^{(j)}) \hat{v}_k^{(j)}}{(\hat{\sigma}_k^{(j)})^2 - (\hat{v}_k^{(j)})^2},$$
$$\hat{\beta}_k^{(j)} = \frac{(1 - \hat{u}_k^{(j)})(\hat{\sigma}_k^{(j)})^2}{(\hat{\sigma}_k^{(j)})^2 - (\hat{v}_k^{(j)})^2};$$

   5) Set $\mathbf{M}^{(j)}(\mathbf{s})$, an $m \times m$ function matrix with elements:
$$M_{kk'}^{(j)}(\mathbf{s}) = -\hat{\phi}_k^{(j)}(s_k) s_{k'}, \ k \neq k'$$
$$= \hat{\alpha}_k^{(j)} s_k + \hat{\beta}_k^{(j)}(2I(|s_k| \leq 1) - 1), \ k = k';$$

   6) Set
$$\hat{\mathbf{l}}^{*(j)}(\mathbf{x}) = vec(\mathbf{M}^{(j)}(\hat{W}^{(j)}\mathbf{x})[\hat{W}^{(j)}]^{-T}),^*$$
     where $vec(M)$ vectorizes $M$;

   7) Set
$$\mathbf{e}_n^{(j)} = \frac{1}{n} \sum_{i=1}^n \hat{\mathbf{l}}^{*(j)}(X^i),$$
$$\Sigma_n^{(j)} = \frac{1}{n} \sum_{i=1}^n \hat{\mathbf{l}}^{*(j)} \hat{\mathbf{l}}^{*(j)T}(X^i);$$

   8) Update $\hat{W}^{(j+1)} = \hat{W}^{(j)} + [\Sigma_n^{(j)}]^{-1} \mathbf{e}_n^{(j)}$.

---

*Here, $\hat{\mathbf{l}}^{*(j)}(\mathbf{x}) = \mathbf{l}^*(\mathbf{x}, W, \hat{\Phi}_W)$ at $W = \hat{W}^{(j)}$, where $\hat{\Phi}_W$ can be identified as an estimate of the "nuisance parameter" $\Phi_W$.

where $\mathbf{\Phi}_W$ is the parameter defined by (2.9) and $\hat{\Phi}_W$ is an estimate of it. We give pseudo code in Table 1 for iteratively solving this equation. The expressions appearing in the pseudo code are developed in the rest of this section.

2.3. *Efficient score function of $W$*. The likelihood function of $X$ under (1.1) can be expressed as

$$p_X(\mathbf{x}, W, r_{1:m}) = |\det(W)| \prod_{i=1}^m r_i(W_i \mathbf{x}).$$

The parameter of interest is $W$, while $r_{1:m}$ are nuisance parameters. For simplicity, assume $E[S_i] = 0$.

Let $\phi_i(s_i) = -\frac{\partial}{\partial s_i} \log r_i(s_i) I(r_i(s_i) > 0)$ be the density score function associated with $r_i$ and define $\Phi$ by $\Phi(\mathbf{s}) = (\phi_1(s_1), \ldots, \phi_m(s_m))^T$, where $\mathbf{s} = (s_1, \ldots, s_m)^T$. Then the score function of $W$, $\dot{\mathbf{l}}_W(\mathbf{x}) \equiv \frac{\partial}{\partial vec(W)} \log(p_X(\mathbf{x}, W, r_{1:m}))$, is equal to

$$\dot{\mathbf{l}}_W(\mathbf{x}) = vec\{(I_{m \times m} - \Phi(\mathbf{s})\mathbf{s}^T)W^{-T}\},$$

where $\mathbf{s} = W\mathbf{x}$ and $I_{m \times m}$ is an $m \times m$ identity matrix. Thus, minimal regularity conditions for efficient estimation are that each $r_i$ should be absolutely continuous, $W$ nonsingular and

$$E[\phi_i(S_i)^2] < \infty \quad \text{and} \quad E[S_i^2] < \infty.$$

Using the devices of tangent space and projection mentioned in Section 2.1 (see calculation details in the Appendix), the efficient score can be expressed as

$$(2.4) \qquad \mathbf{l}^*(\mathbf{x}, W, \Phi) = vec(\mathbf{M}(W\mathbf{x})W^{-T}),$$

where $\mathbf{M}(\mathbf{s})$ is an $m \times m$ function matrix with elements

$$(2.5) \qquad M_{ij}(\mathbf{s}) = -\phi_i(s_i)s_j, \qquad \text{for } 1 \leq i \neq j \leq m,$$

$$(2.6) \qquad M_{ii}(\mathbf{s}) = \alpha_i s_i + \beta_i \kappa(s_i), \qquad \text{for } i = 1, \ldots, m$$

and

$$(2.7) \qquad \alpha_i = -\frac{(1-u_i)v_i}{\sigma_i^2 - v_i^2}, \qquad \beta_i = \frac{(1-u_i)\sigma_i^2}{\sigma_i^2 - v_i^2}, \qquad \sigma_i^2 = E[S_i^2],$$

$$(2.8) \qquad v_i = E[2S_i I(|S_i| \leq 1)], \qquad u_i = E[2S_i \phi_i I(|S_i| \leq 1)].$$

Most of these formulas were derived in [2], but in a different context. We repeat these in our own notation for completeness. By the convolution theorem on semiparametric models (see [5]), the information bound for regular estimators of $W$ is $(E[\mathbf{l}^* \mathbf{l}^{*T}(X, W, \Phi)])^{-1}$. It is obvious that the efficient score function depends on $r_{1:m}$ only through the density score functions $(\phi_1, \ldots, \phi_m)$. Next, we describe how to perform the estimation by using the efficient score function.

2.4. *The ICA estimate.* Let

$$(2.9) \qquad \Phi_W = (\phi_{W_1}, \ldots, \phi_{W_m})^T$$

and assume that a starting estimate $\hat{W}^{(0)}$ is available which is consistent for $W_P$. We shall show how to construct an estimate $\hat{\Phi}_W$ of $\Phi_W$ for $W$ close to $W_P$ and then solve

$$\int \mathbf{l}^*(X, W, \hat{\Phi}_W) \, dP_n = 0$$

to obtain an efficient estimator of $W_P$. Here, $\hat{\Phi}_W$ is a data-dependent function of $W$ and, thus, $\mathbf{l}^*(X, W, \hat{\Phi}_W)$ is an approximation to the efficient score function given by (2.4).

For each $k \in \{1, \ldots, m\}$, choose a sieve for $\hat{\phi}_{W_k}$ as follows. Let $[\underline{b}_{nk}, \overline{b}_{nk}] \subset \mathbb{R}$ be a subset of $supp(r_k)$ containing most of the mass of $r_k$. For an integer $n_k$, set $n_k + 4$ equally spaced points $\{\underline{b}_{nk} + (i-1)\delta_{nk} : 1 \leq i \leq n_k + 4\}$ as knots, where $\delta_{nk}$ depends on $n_k$ through

$$\delta_{nk} = (\overline{b}_{nk} - \underline{b}_{nk})/(n_k + 3),$$

and then construct $n_k$ cubic B-spline basis functions, as in the Appendix. Here, $n_k$ is chosen by cross-validation as described in Section 2.5 below. Denote the basis functions as $\mathbf{B}_n^{(k)} \equiv (B_{n1}^{(k)}, \ldots, B_{nn_k}^{(k)})^T$, where the superscript $(k)$ denotes the association with $S_k$ and the subscript $n$ denotes the dependence on the sample size. Given the random sample $\{W_k X^i : 1 \leq i \leq n\}$ from the density function $f_{W_k}$, the density score function $\phi_{W_k}$ can be estimated by

$$(2.10) \qquad \hat{\phi}_{W_k} = [\gamma_n(W_k)]^T \mathbf{B}_n^{(k)},$$

where $\gamma_n(W_k)$ is given in Table 5 (see Section 2.5 for details). Then define $\hat{\Phi}_W(\mathbf{s}) \equiv (\hat{\phi}_{W_1}(s_1), \ldots, \hat{\phi}_{W_m}(s_m))^T$. To avoid further complications, it is assumed that both $[\underline{b}_{nk}, \overline{b}_{nk}]$ and $n_k$ are fixed using $\hat{W}^{(0)}$. That is, the $n_k + 4$ knot locations are fixed.

Now, replace the efficient score function $\mathbf{l}^*(X, W, \Phi)$ defined in (2.4)–(2.8) by its profile form $\mathbf{l}^*(X, W, \hat{\Phi}_W)$, where $\alpha_i$, $\beta_i$ and $\sigma_i^2$ defined in (2.7) and (2.8) are replaced with plug-in estimates

$$(2.11) \qquad \hat{\alpha}_i = -\frac{(1 - \hat{u}_i)\hat{v}_i}{\hat{\sigma}_i^2 - \hat{v}_i^2}, \qquad \hat{\beta}_i = \frac{(1 - \hat{u}_i)\hat{\sigma}_i^2}{\hat{\sigma}_i^2 - \hat{v}_i^2}, \qquad \hat{\sigma}_i^2 = \int (W_i X)^2 \, dP_n,$$

where $\hat{u}_i = \int_{Y=W_i X} 2Y \hat{\phi}_{W_i}(Y) I(|Y| \leq 1) \, dP_n$ and $\hat{v}_i = \int_{Y=W_i X} 2Y I(|Y| \leq 1) \, dP_n$.

Define

$$(2.12) \quad \mathbf{e}_n(W) = \int \mathbf{l}^*(X, W, \hat{\Phi}_W) \, dP_n \quad \text{and} \quad \mathbf{e}(W) = \int \mathbf{l}^*(X, W, \Phi_W) \, dP$$

and let $\hat{W}$ be a solution of

$$(2.13) \qquad \mathbf{e}_n(W) = 0,$$

if it exists. Let $\hat{\mathbf{l}}^*(\mathbf{x}, W) \equiv \mathbf{l}^*(\mathbf{x}, W, \hat{\Phi}_W)$ and $\dot{\mathbf{e}}_n(W) \equiv \frac{\partial}{\partial vec(W)} \mathbf{e}_n(W)$. Note that if $\hat{W} \to W_P$, then $-\dot{\mathbf{e}}_n(\hat{W})$ and $\int \hat{\mathbf{l}}^* \hat{\mathbf{l}}^{*T}(X, \hat{W}) \, dP_n$ have the same limit,

$$-\frac{\partial \mathbf{e}(W)}{\partial vec(W)}\bigg|_{W_P} = E[\mathbf{l}^* \mathbf{l}^{*T}(X, W_P, \Phi_P)],$$

with probability converging to 1, as demonstrated later in Section 5. The final estimator $\hat{W}$ is defined as the limiting value of the following approximate Newton–Raphson iteration:

$$(2.14) \quad \hat{W}^{(j+1)} = \hat{W}^{(j)} + \left[ \int \hat{\mathbf{I}}^* \hat{\mathbf{I}}^{*T}(X, \hat{W}^{(j)}) \, dP_n \right]^{-1} \mathbf{e}_n(\hat{W}^{(j)}), \qquad j = 0, 1, \ldots.$$

We shall show that this limit exists with probability tending to 1. Note that this method does not require the calculation of the Hessian matrix $\dot{\mathbf{e}}_n(W)$. The convergence and asymptotic properties of (2.14) are developed in Section 3. Call $\hat{W} \equiv \hat{W}^{(\infty)}$ defined by (2.14) the *EFFICA estimate*. This is summarized in Table 1 and will be used for the simulation in Section 4. The density score estimation, as well as how to choose the number of knots by cross-validation (mentioned above), is provided in the next subsection.

2.5. *Estimating a density score function by B-spline approximations.* Let $\phi = -r'/r$ be the density score associated with a univariate PDF $r$. Let $\mathcal{G}$ be a linear space with differentiable basis functions $\mathbf{B} = (B_1, \ldots, B_N)^T$ such that each $B_i r$ vanishes at infinity. An estimator of $\phi$ in $\mathcal{G}$ can then be obtained by minimizing with respect to $\gamma \in R^N$ the mean square error

$$c(\gamma) = \int_R (\phi(s) - \gamma^T \mathbf{B}(s))^2 r(s) \, ds.$$

Using integration by parts, we obtain

$$c(\gamma) = \gamma^T E_r[\mathbf{B}\mathbf{B}^T]\gamma - 2\gamma^T E_r[\mathbf{B}'] + E_r[\phi^2],$$

where $\mathbf{B}'$ is the derivative of $\mathbf{B}$ and $E_r$ indicates expectation under $r$. Thus, the optimal $\gamma$ is $\gamma_\phi = (E_r[\mathbf{B}\mathbf{B}^T])^{-1} E_r[\mathbf{B}']$ and the best approximation of $\phi$ in $\mathcal{G}$, in the sense of mean square error, is $\phi_{\mathcal{G}} = \gamma_\phi^T \mathbf{B}$. This method was proposed by Jin [17] as a variant of Cox's [10] penalized estimators. Given a random sample of size $n$ from the density function $r$, $\gamma_\phi$ can be estimated by combinations of empirical moments. So, a natural estimator of $\phi$ is given by

$$(2.15) \qquad \hat{\phi}_{\mathcal{G}} = \hat{\gamma}_\phi^T \mathbf{B}, \qquad \text{where } \hat{\gamma}_\phi^T = (\hat{E}_r[\mathbf{B}\mathbf{B}^T])^{-1} \hat{E}_r[\mathbf{B}'],$$

where $\hat{E}_r$ is empirical expectation corresponding to $E_r$.

B-spline basis functions are popular choices for $\mathcal{G}$. In general, the support of $r$ is unknown and we need to choose a working interval $[\underline{b}_n, \overline{b}_n]$, in which knots are distributed, for the construction of the basis functions. The basic rule for adaptation is that $[\underline{b}_n, \overline{b}_n] \to supp(r)$ very slowly as $n \to \infty$. Here, $\underline{b}_n$ and $\overline{b}_n$ are selected as $\alpha_n$ and $1 - \alpha_n$ empirical quantiles where $\alpha_n \to 0$. The number of basis functions, say $N$, is an additional empirical smoothing parameter. One can use cross-validation to choose $N$ as follows:

1. Split the sample randomly into two halves, say $\mathcal{I}_1$ and $\mathcal{I}_2$;

2. For $N = 1, 2, \ldots$, use $\mathcal{I}_1$ to estimate $\gamma_\phi \in \mathbb{R}^N$ by (2.15), say $\hat{\gamma}_\phi(\mathcal{I}_1)$, and use $\mathcal{I}_2$ to evaluate $c(\hat{\gamma}_\phi(\mathcal{I}_1))$ empirically, but omitting the last term $E_r[\phi^2]$, say $\hat{c}_{\mathcal{I}_2|\mathcal{I}_1}(N)$. Similarly calculate $\hat{c}_{\mathcal{I}_1|\mathcal{I}_2}(N)$;

3. Select $N$ as the largest value such that $\frac{1}{2}\{\hat{c}_{\mathcal{I}_2|\mathcal{I}_1}(N) + \hat{c}_{\mathcal{I}_1|\mathcal{I}_2}(N)\}$ strictly decreases until $N$.

Jin [17] used a similar method in the i.i.d. case we have discussed, proved its validity and showed that $N = O(n^\delta)$ under weak smoothness assumptions, where $\delta \in (0, 1/6)$ depends on tail properties of $r$.

## 3. Asymptotic properties.

We are given $\hat{W}^{(0)}$ (e.g., the PCFICA estimate) such that for some $\varepsilon_n > 0$,

$$(3.1) \qquad P(\|\hat{W}^{(0)} - W_P\|_F \le \varepsilon_n) \to 1$$

as $n \to \infty$, where $\varepsilon_n$ satisfies $\varepsilon_n \to 0$ and $\sqrt{n}\varepsilon_n \to \infty$. Let

$$(3.2) \qquad \Omega_n = \{W \in \mathbb{R}^{m \times m} : \|W - W_P\|_F < \varepsilon_n\}.$$

Define $\phi_{W_k, n}(x) \equiv \phi_{W_k}(x) I(x \in [\underline{b}_{nk}, \overline{b}_{nk}])$ and consider the following conditions for $1 \le k, i, j \le m$:

C1: $W_P$ is nonsingular.

C2: $E[S_k] = 0$, $E[S_k^2] < \infty$, $med(|S_k|) = 1$ and $E(\phi_k(S_k))^2 < \infty$.

C3: $|r_k|_\infty < \infty$, $|r_k'|_\infty < \infty$, $\sup_{t \in \mathcal{R}} |tr_k'(t)| < \infty$.

C4: The uniform law of large numbers (ULLN) holds for $\{\phi_{W_k}(W_k X) X_i : W \in \Omega_n\}$, $\{\phi_{W_k}'(W_k X) X_i^2 : W \in \Omega_n\}$ and for $\{\phi_{W_k}'(W_k X) W_i X X_j : W \in \Omega_n\}$.

C5: For some positive $c_1, c_2$, $r_k(t) \ge c_1 \delta_{nk}$ if $t \in [\underline{b}_{nk}, \overline{b}_{nk}]$, otherwise $r_k(t) \le c_2 \delta_{nk}$.

C6: $\sup_{W \in \Omega_n} |\phi_{W_k, n}|_\infty \delta_{nk} = O(1)$ and $\sup_{W \in \Omega_n} |\phi_{W_k, n}'''|_\infty \delta_{nk} = o(1)$.

C7: $\varepsilon_n \delta_{nk}^{-\frac{11}{2}} (\overline{b}_{nk} - \underline{b}_{nk}) = o(1)$.

[Note: ULLN holds for $\mathcal{G}_n$ iff $\sup_{g \in \mathcal{G}_n} |\int g(X) d(P_n - P)| = o_p(1)$; see, e.g., [24].]

Conditions C1–C3 are simplified regularity conditions. C1 and the finite moments in C2 are among the minimal regularity conditions for considering efficiency, as mentioned in Section 2.3. Setting the absolute median to unity in C2 is a simple and minimal condition to make the scales of the unmixing matrix identifiable [9]. It should be clear that the zero mean assumption in C2 is in no way crucial to the general argument as the mean can be estimated adaptively, but it serves to keep algebraic complication to a minimum. C3 assumes some smoothness on the density score function $\phi_k$ for each hidden component, which is needed if it is to be well approximated by B-splines.

Conditions C4–C7 are technical conditions that we believe are far from necessary, but they are reasonably easy to check and enable construction of a more compact proof. As an easy example, if $|\phi_k|_\infty < \infty$ and $|r_k''/r_k|_\infty < \infty$ for $k = 1, \ldots, m$, then by (A.1) in the Appendix, $\sup_{W \in \Omega_n} |\phi_{W_k}|_\infty < \infty$ and by (A.2), $\sup_{\Omega_n} |\phi_{W_k}'|_\infty < \infty$. Thus C4 holds. C5 and C6 require that the tail of $r_k$ be not too wiggly. C6 also implies that $\delta_{nk} \to 0$. C7 requires that the initial value be reasonably close to the truth and that the domain and the number of knots of the B-splines [i.e., $n_k = (\overline{b}_{nk} - \underline{b}_{nk})\delta_{nk}^{-1} - 3$] do not grow so quickly that we lose control of the approximation to $\Phi_W$.

Here is the main theorem:

THEOREM 3.1. _In the ICA model_ (1.1), _if_ (3.1) _and_ C1–C7 _hold for_ $i, j, k = 1, \ldots, m$, $i \neq k$ _and_ $j \neq k$, _then with probability converging to_ 1, _the algorithm_ (2.14) _has a limit_ $\hat{W}^{(\infty)}$ _and_

$$(3.3) \quad \sqrt{n}\,vec(\hat{W}^{(\infty)} - W_P) = \mathbf{I}_{\text{eff}}^{-1} \sqrt{n} \int \mathbf{l}^*(X, W_P, \Phi_P)\,dP_n + o_P(1),$$

_where_ $\mathbf{I}_{\text{eff}} = \int \mathbf{l}^* \mathbf{l}^{*T}(X, W_P, \Phi_P)\,dP$. _That is,_ $\hat{W}^{(\infty)}$ _is Fisher efficient. Further,_

$$\sqrt{n}\,vec(\hat{W}^{(\infty)} W_P^{-1} - I_{m \times m}) \to_d \mathcal{N}(\mathbf{0}, \bar{\mathbf{I}}_{\text{eff}}^{-1}),$$

_where_ $\bar{\mathbf{I}}_{\text{eff}} = \int vec(\mathbf{M})vec(\mathbf{M})^T\,dP$ _does not depend on_ $W_P$ _and_ $\mathbf{M}$ _is given by_ (2.5)–(2.8), _with_ $\mathbf{s}$ _replaced by_ $S$.

The proof of Theorem 3.1 is given in later sections and the Appendix.

## 4. Numerical studies and some computational issues.

Two groups of experiments are implemented to test the empirical performance of EFFICA. Data are generated from known source distributions listed in Table 2 with a known mixing matrix $W_P^{-1}$. The boundaries for B-spline approximation of the density score functions are taken as $\underline{b}_{nk} = \max(q_n(0), q_n(0.01) - \Delta_n)$ and $\overline{b}_{nk} = \min(q_n(1), q_n(0.99) + \Delta_n)$, where $q_n(\cdot)$ denotes the empirical quantile and $\Delta_n = c \cdot \sqrt{\log \log n}$. We used $c = 5$ in the simulation. The number of knots is key for EFFICA and is chosen by the cross-validation method described in Section 2.5.

In the first group of experiments, two hidden components are used and $W_P = [2, 1; 2, 3]$. The two components in the first twelve experiments are i.i.d from one of the distributions [1]–[12], and the two components in experiments 13–15 are from different distributions as specified in cases [13]–[15] of Table 2. Each experiment has been replicated 400 times with $n = 1000$.

In the second group of experiments, the number of hidden components is increased to $m = 4$, 8 and 12, $m$ hidden components are chosen with distributions of $[0], [1], \ldots, [m - 1]$ in Table 2 and $W_P = I_{m \times m}$. The experiments are replicated 100, 100 and 50 times for $m = 4$, 8 and 12, respectively, with $n = 4000$.

Comparisons are made with five existing ICA algorithms: the FastICA algorithm with the options of "symmetric" and "tanh" [15] which is equivalent to quasi-ML by using a tanh distribution for each hidden source (note: the FastICA code uses logistic for tanh), the JadeICA algorithm [7], the extended infomax algorithm [19], the KernelICA-Kgv algorithm [3] and the PCFICA algorithm. The PCFICA's estimate is used as the initial value for both EFFICA and KernelICA-Kgv. Due to the existence of multiple local solutions, PCFICA uses three starting values, one from FastICA and the others random. The performance of each algorithm is measured by both the Frobenius error, that is, $d_F(\hat{W}, W_P) = \|\hat{W}W_P^{-1} - I_{m\times m}\|_F$ after suitable rescaling and permutation on rows of both $\hat{W}$ and $W_P$, and the so-called *Amari error* $d_A(\hat{W}, W_P)$ (e.g., [3]),

$$d_A(V, W) = \frac{1}{2m}\sum_{i=1}^{m}\left(\frac{\sum_{j=1}^{m}|a_{ij}|}{\max_j |a_{ij}|} - 1\right) + \frac{1}{2m}\sum_{j=1}^{m}\left(\frac{\sum_{i=1}^{m}|a_{ij}|}{\max_i |a_{ij}|} - 1\right),$$

where $V, W$ are rescaled into $\bar{V}, \bar{W}$ such that each row of $\bar{V}$ and $\bar{W}$ has norm 1 and $a_{ij} = (\bar{V}\bar{W}^{-1})_{ij}$. The Amari error lies in $[0, m-1]$, is invariant under permutation and scaling of the rows of $V$ and $W$ and is equal to zero if and only if $V$ and $W$ represent the same row components.

For each experiment in the first group of simulations with $T = 400$ replications, Table 3 reports the average Amari error and square root of the mean square error $\sqrt{MSE}$ with

$$MSE = \frac{1}{T}\sum_{i=1}^{T}(d_F^{(i)})^2,$$

where $d_F^{(i)}$ denotes the Frobenius error for the $i$th replication. For the second group of simulations, Figure 1 shows the boxplots of the Amari errors and Table 4 reports $\sqrt{MSE}$.

From the simulation results, in some cases, some parametric ICA algorithms work very well and even outperform EFFICA. For example, FastICA

TABLE 2
*Source distributions used in the simulations*

| | |
|---|---|
| [0]. N(0,1) | [8]. exp(1) + U(0,1) |
| [1]. exp(1) | [9]. mixture exp. |
| [2]. t(3) | [10]. mixture of exp. and normal |
| [3]. lognormal(1, 1) | [11]. mixture Gaussians: multimodal |
| [4]. t(5) | [12]. mixture Gaussians: unimodal |
| [5]. logistic(0, 1) | [13]. exp(1) vs normal(0,1) |
| [6]. Weibull(3, 1) | [14]. lognormal(1, 1) vs normal(0, 1) |
| [7]. exp(10) + normal(0, 1) | [15]. Weibull(3, 1) vs exp(1) |

Table 3
*1000 × mean Amari errors (and $1000 \times \sqrt{MSE}$ in brackets) for six ICA methods using $m = 2$ sources and sample size $n = 1000$. Source distributions for row $k$ are given by $[k]$ in Table 1. The bold numbers represent the best performance according to each experiment*

| pdf | Fast | Jade | ExtImax | Pcf | Kgv | EFFICA |
|-----|------|------|---------|-----|-----|--------|
| 1 | 37 | 39 | 34 | 18 | 14 | **7** |
|   | (89) | (66) | (57) | (31) | (24) | **(11)** |
| 2 | 36 | 36 | **24** | 35 | 33 | 29 |
|   | (231) | (68) | (61) | (61) | (55) | **(52)** |
| 3 | 33 | 31 | 19 | 16 | 14 | **5** |
|   | (243) | (69) | (33) | (27) | (24) | **(8)** |
| 4 | **39** | 50 | 41 | 60 | 61 | 60 |
|   | (99) | **(86)** | (112) | (100) | (102) | (110) |
| 5 | **71** | 85 | 87 | 109 | 99 | 128 |
|   | (194) | **(153)** | (232) | (192) | (170) | (253) |
| 6 | 42 | 43 | 32 | 18 | 15 | **7** |
|   | (188) | (83) | (57) | (30) | (24) | **(11)** |
| 7 | 43 | 41 | 35 | 18 | 15 | **9** |
|   | (205) | (72) | (96) | (31) | (25) | **(16)** |
| 8 | 36 | 44 | 35 | 21 | 19 | **17** |
|   | (99) | (96) | (64) | (35) | (31) | **(28)** |
| 9 | 35 | 37 | 24 | 16 | 14 | **4** |
|   | (212) | (83) | (41) | (28) | (24) | **(7)** |
| 10 | 46 | 59 | 39 | 44 | **30** | 47 |
|    | (209) | (103) | (66) | (74) | **(49)** | (105) |
| 11 | 28 | 33 | 27 | 29 | **25** | 25 |
|    | (47) | (54) | (45) | (48) | **(41)** | (42) |
| 12 | 50 | 49 | 44 | 44 | **39** | 78 |
|    | (184) | (82) | (78) | (74) | **(66)** | (264) |
| 13 | 65 | 52 | 185 | 24 | 19 | **16** |
|    | (164) | (89) | (355) | (42) | (35) | **(31)** |
| 14 | 35 | 45 | 91 | 20 | 14 | **11** |
|    | (109) | (89) | (188) | (35) | (24) | **(20)** |
| 15 | 69 | 72 | 57 | 32 | 27 | **11** |
|    | (192) | (184) | (136) | (69) | (58) | **(25)** |

works best in case 5 where hidden sources have logistic distributions. This is not surprising as we have pointed out in Section 2.1 that a simple quasi-MLE can outperform an efficient estimator when the value of the nuisance parameter used by the quasi-MLE is close to the truth. But, in most experiments, the parametric methods (FastICA, JADE, ExtImax) behave worse than the nonparametric methods (PCFICA, Kgv, EFFICA) and EFFICA has both the smallest Amari errors and smallest Frobenius errors, while Kgv, which we conjecture can be efficient after appropriate regularization, is the best in the cases of mixture Gaussians. The three nonparametric ICA al-gorithms require heavier computation, but their performance is better than the parametric methods.

TABLE 4
$10 \times \sqrt{MSE}$ for ICA algorithms with the same simulations as in Figure 1

| Case | Fast | Jade | ExtImax | PCF | Kgv | EFFICA |
|------|------|------|---------|-----|-----|--------|
| $m = 4$ | 0.82 | 1.31 | 2.71 | 0.60 | 0.51 | 0.45 |
| $m = 8$ | 7.0 | 8.3 | 11.2 | 5.4 | 4.3 | 3.6 |
| $m = 12$ | 9.1 | 11.2 | 13.1 | 7.5 | 8.0 | 8.7 |

All of the ICA algorithms used in the simulation except EFFICA are based on contrast functions which empirically measure the dependence or nongaussianity among $\{W_1 X, \ldots, W_m X\}$ and, thus, they are invariant with respect to the choice of $W_P$ for both error metrics $d_F(\hat{W}, W_P)$ and $d_A(\hat{W}, W_P)$. We note that prewhitening, which is used for data preprocessing by these algorithms, can reduce such invariance, although it does not cause inconsistency [8]. Theorem 3.1 implies that EFFICA is asymptotically invariant with respect to $W_P$. Figure 2 compares $m = 8$ with two different unmixing matrices $W_P = I_{m \times m}$ and $W_P = I_{m \times m} + V$, where $V_{jk} = j/m^2 + (k-1)/m$ for $1 \le j, k \le m$. We performed many other simulations with different $W_P$ and obtained similar results. We observe that the Frobenius error boxplots do change somewhat with different $W_P$, but EFFICA is more robust than other ICA algorithms. We believe that the main reasons are (i) none of the ICA algorithms are convex and, thus, may suffer from local solutions and (ii) EFFICA does not use prewhitening for preprocessing, while others do.

**5. Proof of Theorem 3.1.** In this section, we prove Theorem 3.1. Note that solving (2.13) can be viewed as a generalized M-estimator (GM-estimator). The existence/uniqueness, convergence and asymptotic linearity of GM-estimators have been studied in [5] (the Iteration Theorem in Appendix A.10.2, page 517). The idea of our proof is to use the Iteration Theorem.

Suppose that $M_n(\theta, P_n)$ is a functional of $\theta \in \Omega$ (a subset of a finite Euclidean space) and $P_n$, but is not necessarily linear in $P_n$. The subscript $n$ in $M_n$ allows the existence of a possible smoothing or sieve parameter dependent on $n$. The zero of $M_n(\theta, P_n)$ w.r.t $\theta$ is called a *generalized M-estimator*. Let $M(\theta, P) \equiv M_\infty(\theta, P)$. We review the conditions for the Iteration Theorem.

[GM1] $\theta_P \in \Omega$ is the unique solution of $M(\theta, P) = 0$ in $\Omega$.

[GM2] $M_n(\theta_P, P_n) = \int \psi_{\theta_P}(X) \, dP_n + o_p(n^{-1/2})$ for some $\psi_{\theta_P} \in L_2(P)$.

[GM3] $M(\theta, P)$ is differentiable w.r.t $\theta$ in a neighborhood of $\theta_P$ and $\frac{\partial M(\theta_P, P)}{\partial \theta}$ is nonsingular.

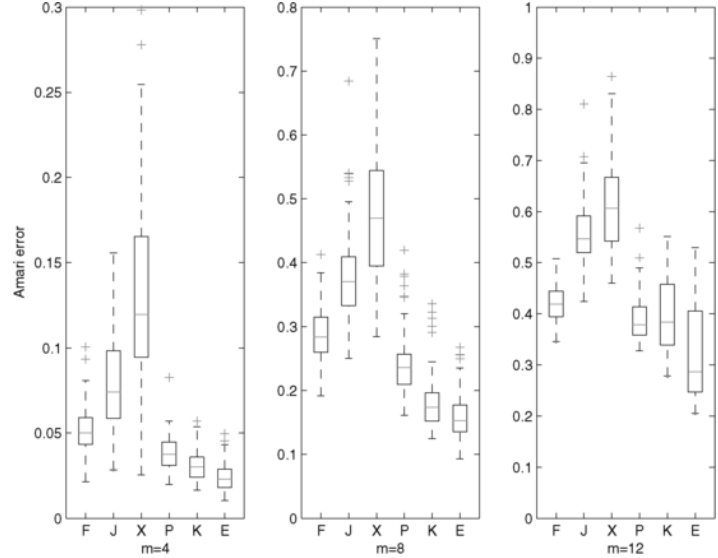

Fig. 1. *Boxplots of Amari errors for ICA algorithms: in each case (left, middle, right), m hidden components are generated from pdfs $[0], \ldots, [m-1]$ in Table 1. The X-labels represent ICA algorithms: F-FastICA, J-JadeICA, X-extended infomax, P-PCFICA, K-Kgv, E-EFFICA. The sample sizes are 4000 for all the experiments and the replication times are 100, 100, 50 for $m = 4$, $m = 8$ and $m = 12$, respectively.*

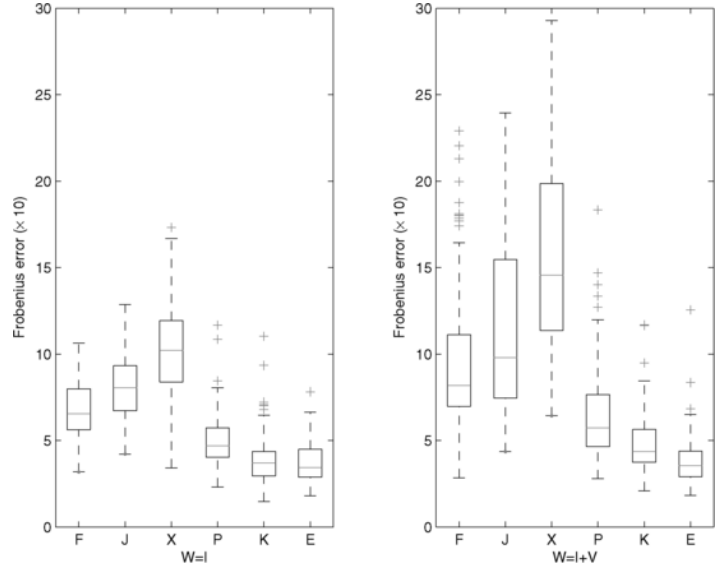

Fig. 2. *Frobenius errors (×10) with $m = 8$ and $n = 4000$, where the left panel is for $W = I_{m \times m}$ and the right panel is for $W = I_{m \times m} + V$, with $V_{jk} = j/m^2 + (k-1)/m$. Each experiment is replicated 100 times.*

For our efficient score equation $M_n(\theta, P_n) = \mathbf{e}_n(W)$ defined in (2.12), [5] condition [U] becomes

[U]  $\sup_{W \in \Omega_n} |\dot{\mathbf{e}}_n(W) - \dot{\mathbf{e}}(W_P)| = o_P(1)$.

THEOREM 5.1 [5].  *Suppose* [GM1], [GM2] *and* [GM3] *hold with* $M_n(\theta, P_n) = \mathbf{e}_n(W)$ *and that* [U] *holds. If the starting point satisfies* $P(|\hat{W}^{(0)} - W_P| < \varepsilon_n) \to 1$, *then with probability converging to* 1, $\mathbf{e}_n(W)$ *in* (2.13) *has a unique root* $\hat{W}^{(\infty)}$ *which is also the limit of the sequence defined by* (2.14), *except with* $\int \hat{\mathbf{l}}^* \hat{\mathbf{l}}^{*T}(X, \hat{W}^{(j)}) \, dP_n$ *replaced by* $-\dot{\mathbf{e}}_n(\hat{W}^{(j)})$, *and* $\hat{W}^{(\infty)}$ *is asymptotically linear with the influence function* $-[\dot{\mathbf{e}}(W_P)]^{-1} \mathbf{l}^*(., W_P, \Phi_P)$.

Theorem 5.1 is called the *Iteration Theorem* in [5]. Note that the sequence limit defined by the Iteration Theorem uses the exact Newton–Raphson, whereas we use an approximate Newton–Raphson, as in (2.14). To make up the difference, we need the following condition [V] which is verified by Proposition 6.4 in Section 6:

[V]  $\sup_{W \in \Omega_n} |\int \hat{\mathbf{l}}^* \hat{\mathbf{l}}^{*T}(X, W) \, dP_n - \int \mathbf{l}^* \mathbf{l}^{*T}(X, W_P, \Phi_{W_P}) \, dP| = o_P(1)$.

Theorem 3.1 can now be proved as follows:

PROOF OF THEOREM 3.1.  It is obvious that [GM1] holds under the conditions of Theorem 3.1 as it is the efficient score function. [GM2], [GM3] and [U] are verified by Propositions 6.1, 6.2 and 6.3 below, respectively. Thus, the conclusion of the above Iteration Theorem applies here. By Proposition 6.4, the condition [V] holds. Further, by Proposition 6.2,

$$\dot{\mathbf{e}}(W_P) = -E[\mathbf{l}^* \mathbf{l}^{*T}(X, W_P, \Phi_P)].$$

Thus, we have

$$\sup_{W \in \Omega_n} \left| \dot{\mathbf{e}}_n(W) + \int \hat{\mathbf{l}}^* \hat{\mathbf{l}}^{*T}(x, W) \, dP_n \right| = o_P(1).$$

Then following the contraction arguments of [5] (pages 317–319), the iteration given in (2.14) has the same limit as that replacing $\int \hat{\mathbf{l}}^* \hat{\mathbf{l}}^{*T}(X, \hat{W}^{(j)}) \, dP_n$ by $-\dot{\mathbf{e}}_n(\hat{W}^{(j)})$, with probability converging to 1. Thus, (3.3) holds. The second result follows from (3.3) directly by using (2.4) and the devices of Kronecker product and *vec* operator.  □

**6. Propositions 6.1–6.4.** This section verifies conditions [GM2], [GM3], [U] and [V]. For convenience, we list all the notation used in the following proofs in Table 5, for $k \in \{1, \ldots, m\}$ and $W \in \Omega_n$. Note that all of the lemmas used in this section are given in the Appendix. For simplicity of notation, we will often write $\delta_{nk}$ as $\delta_n$.

TABLE 5
*List of all notation used in the proof*

| | |
|---|---|
| $P, P_n$ | population, empirical law of $X$ |
| $W, W_k, W_{ij}$ | $m \times m$ matrix, $k$th row, $(i,j)$th element |
| $W_P, W_{Pk}, W_{Pij}$ | unmixing matrix, $k$th row, $(i,j)$th entry |
| $r_k$ | PDF of $S_k$ |
| $\phi_k = -r'_k/r_k$ | density score function for $S_k$ |
| $\Phi_P = (\phi_1, \ldots, \phi_m)^T$ | function vector |
| $f_{W_k}$ | PDF of $W_k X$ ($f_{W_{Pk}} \equiv r_k$) |
| $\phi_{W_k} = -f'_{W_k}/f_{W_k}$ | score function of $W_k X$ ($\phi_{W_{Pk}} \equiv \phi_k$) |
| $\phi_{k,n}, \phi_{W_k,n}$ | truncation of $\phi_k, \phi_{W_k}$ on $[\underline{b}_{nk}, \overline{b}_{nk}]$ |
| $\Phi_W = (\phi_{W_1}, \ldots, \phi_{W_m})^T$ | function vector |
| $\hat{\Phi}_W = (\hat{\phi}_{W_1}, \ldots, \hat{\phi}_{W_m})^T$ | function vector |
| $\mathbf{l}^*(X, W, \Phi)$ | efficient score function defined by (2.4) |
| $\mathbf{e}(W) = \int \mathbf{l}^*(X, W, \Phi_W)\, dP$ | expectation |
| $\mathbf{e}_n(W) = \int \mathbf{l}^*(X, W, \hat{\Phi}_W)\, dP_n$ | empirical expectation |
| $\mathbf{B}_n^{(k)} = (B_{n1}^{(k)}, \ldots, B_{nn_k}^{(k)})^T$ | B-splines defined on $[\underline{b}_{nk}, \overline{b}_{nk}]$ |
| $A_n(W_k) = \int \mathbf{B}_n^{(k)} \mathbf{B}_n^{(k)T}(W_k X)\, dP_n$ | served in coefficients of $\hat{\phi}_{W_k}$ |
| $D_n(W_k) = \int (\mathbf{B}_n^{(k)})'(W_k X)\, dP_n$ | served in coefficients of $\hat{\phi}_{W_k}$ |
| $\gamma_n(W_k) = A_n(W_k)^{-1} D_n(W_k)$ | served as coefficients of $\hat{\phi}_{W_k}$ |
| $A(W_k) = \int \mathbf{B}_n^{(k)} \mathbf{B}_n^{(k)T}(W_k X)\, dP$ | served in coefficients of $\overline{\hat{\phi}}_{W_k}$ |
| $D(W_k) = \int (\mathbf{B}_n^{(k)})'(W_k X)\, dP$ | served in coefficients of $\overline{\hat{\phi}}_{W_k}$ |
| $\gamma(W_k) = A(W_k)^{-1} D(W_k)$ | served as coefficients of $\overline{\hat{\phi}}_{W_k}$ |
| $\mathcal{G}_n^{(k)} = \{a^T \mathbf{B}_n^{(k)} : a \in \mathcal{R}^{n_k}\}$ | closed linear span of B-splines |
| $\hat{\phi}_{W_k} = \gamma_n(W_k)^T \mathbf{B}_n^{(k)}$ | estimator of $\phi_{W_k}$ in $\mathcal{G}_n^{(k)}$ defined by (2.10) |
| $\overline{\hat{\phi}}_{W_k} = \gamma(W_k)^T \mathbf{B}_n^{(k)}$ | estimator of $\phi_{W_k}$ in $\mathcal{G}_n^{(k)}$ defined by (A.3) |

PROPOSITION 6.1. *Under the conditions of Theorem* 3.1, *we have*

$$\mathbf{e}_n(W_P) = \int \mathbf{l}^*(X, W_P, \Phi_P)\, dP_n + o_P(n^{-1/2}).$$

PROOF. Recall the definition of $\mathbf{e}_n(W)$ given by (2.12) and that of $\mathbf{l}^*(\mathbf{x}, W, \Phi)$ given by (2.4)–(2.8). It is sufficient to show that for $1 \le i \ne j \le m$, $\hat{\alpha}_i - \alpha_i = o_P(1)$ and $\hat{\beta}_i - \beta_i = o_P(1)$, where $(\alpha_i, \beta_i)$ and $(\hat{\alpha}_i, \hat{\beta}_i)$ are defined in (2.7) and (2.11), respectively, and

(6.1) $$\int \hat{\phi}_{W_{Pi}}(S_i) S_j \, dP_n = \int \phi_{W_{Pi}}(S_i) S_j \, dP_n + o_P(n^{-1/2}),$$

where $S_i = W_{Pi} X$, $S_j = W_{Pj} X$.

The first two are not hard verify with the law of large numbers and Lemma A.7. Here, we will just show (6.1). Observe that

$$\left| \int \hat{\phi}_{W_{Pi}}(S_i) S_j \, dP_n - \int \phi_{W_{Pi}}(S_i) S_j \, dP_n \right| = \left| \int [\hat{\phi}_{W_{Pi}}(S_i) - \overline{\hat{\phi}}_{W_{Pi}}(S_i)] S_j \, dP_n \right|$$

$$+ \left| \int [\overline{\dot{\phi}}_{W_{Pi}}(S_i) - \phi_{i,n}(S_i)] S_j \, dP_n \right|$$

$$+ \left| \int (\phi_i(S_i) - \phi_{i,n}(S_i)) S_j \, dP_n \right|$$

$$= [1] + [2] + [3].$$

In the following, we will show that all of [1], [2] and [3] are $o_P(n^{-1/2})$.

First, by Lemma A.2(7) and Lemma A.3,

$$[1] = \left| \int (\gamma_n(W_{Pi}) - \gamma(W_{Pi}))^T \mathbf{B}_n^{(i)}(S_i) S_j \, dP_n \right|$$

$$\leq \| \gamma_n(W_{Pi}) - \gamma(W_{Pi}) \|_2 \left\| \int \mathbf{B}_n^{(i)}(S_i) S_j \, dP_n \right\|_2$$

$$= o_P(1) O_P(n^{-1/2}).$$

Further, $E([2])^2 = \frac{1}{n} E(\overline{\dot{\phi}}_{W_{Pi}}(S_i) - \phi_{W_{Pi},n}(S_i))^2 E(S_j^2)$. By Lemma A.6, $|\overline{\dot{\phi}}_{W_{Pi}} - \phi_{W_{Pi},n}|_\infty \leq c\delta_n^2 |\phi_{W_{Pi},n}'''|_\infty$. Thus, by C6,

$$[2] = n^{-1/2} \delta_n^2 |\phi_{W_{Pi},n}'''|_\infty O_P(1) = o_P(n^{-1/2}).$$

For [3], since $P(S_i \notin [\underline{b}_{ni}, \overline{b}_{ni}]) \to 0$, we have

$$E([3])^2 = \frac{1}{n} E(\phi_i(S_i)^2 I(S_i \notin [\underline{b}_{ni}, \overline{b}_{ni}])) E(S_j^2) = o\left(\frac{1}{n}\right).$$

So $[3] = o_P(n^{-1/2})$.  □

PROPOSITION 6.2.   *Under conditions* C1, C2 *and* C4 *of Theorem* 3.1, $\mathbf{e}(W)$ *is differentiable w.r.t.* $W$ *in a neighborhood of* $W_P$ *and*

$$\dot{\mathbf{e}}(W_P) = -E\{\mathbf{l}^* \mathbf{l}^{*T}(X, W_P, \Phi_P)\}$$

*is nonsingular.*

PROOF.   Let $T_w(\cdot) = \frac{\partial}{\partial w}\{\phi_w(\cdot)\}$ for any nonzero row vector $w \in \mathbb{R}^m$. By (A.1) in the Appendix, after exchanging the order of differentiation and integration, we have $E[T_w(wX)] = 0$. Then by (2.5), we have for $k = 1, \ldots, m$,

$$E\left\{\frac{\partial}{\partial W_k}\{\mathbf{l}^*(X, W, \Phi_W)\}_{W_P}\right\} = E\left\{\frac{\partial}{\partial W_k}\{\mathbf{l}^*(X, W, \Phi_P)\}_{W_P}\right\}.$$

Since the left-hand side of the above is $\dot{\mathbf{e}}(W_P)$, by Lemma A.8, the right-hand side is equal to

$$(6.2) \qquad \dot{\mathbf{e}}(W_P) = -E\{\mathbf{l}^* \mathbf{l}^{*T}(X, W_P, \Phi_P)\}.$$

Note that the elements of $\mathbf{M}$ in (2.5)–(2.6) are linearly independent, and that this is also true for the elements of $\mathbf{l}^*(., W_P, \Phi_P)$. Thus, $\dot{\mathbf{e}}(W_P)$ must be nonsingular.  □

PROPOSITION 6.3. *Under the conditions of Theorem* 3.1, *for* $i, j, k = 1, \ldots, m$, *we have*

$$\tag{6.3} \sup_{\Omega_n} \left| \int \hat{\phi}_{W_i}(W_i X) X_k \, dP_n(X) - \int \phi_{Pi}(W_{Pi} X) X_k \, dP \right| = o_P(1)$$

*and for* $i \neq j$,

$$\sup_{\Omega_n} \left| \int \frac{\partial}{\partial W_i} \{\hat{\phi}_{W_i}(W_i X)\} W_j X \, dP_n - \int \frac{\partial}{\partial W_i} \{\phi_{Pi}(W_{Pi} X)\} W_{Pj} X \, dP \right|$$
$$\tag{6.4} \qquad = o_P(1).$$

*Thus, condition* [U] *holds.*

PROOF. We omit the superscript $(i)$ in $B_n^{(i)}$ henceforth. By the Cauchy–Schwarz inequality,

$$\left\| \int \mathbf{B}_n^{(i)}(W_i X) X_k \, dP_n \right\|_2^2 \leq \int |X_k|^2 \, dP_n \int \sum_{l=1}^{n_i} [B_{nl}(W_i X)]^2 \, dP_n.$$

Since $\sum_{l=1}^{n_i} [B_{nl}(W_i X)]^2 < 1$ by property III of B-splines in the Appendix, we have

$$\tag{6.5} \sup_{\Omega_n} \left\| \int \mathbf{B}_n(W_i X) X_k \, dP_n \right\|_2 = O_P(1)$$

and by Lemma A.2(7), $\sup_{\Omega_n} \|\gamma_n(W_k) - \gamma(W_k)\|_2 = o_P(1)$, so

$$\sup_{\Omega_n} \left| \int \hat{\phi}_{W_i}(W_i X) X_k \, dP_n(X) - \int \overline{\hat{\phi}}_{W_i}(W_i X) X_k \, dP_n \right|$$
$$= \sup_{\Omega_n} \left| (\gamma_n(W_k) - \gamma(W_k))^T \int \mathbf{B}_n(W_i X) X_k \, dP_n \right|$$
$$\tag{6.6}$$
$$\leq \sup_{\Omega_n} \|\gamma_n(W_k) - \gamma(W_k)\|_2 \sup_{\Omega_n} \left\| \int \mathbf{B}_n(W_i X) X_k \, dP_n \right\|_2$$
$$= o_P(1) O_P(1).$$

Further, by Lemma A.6, $\sup_{\Omega_n} |\overline{\hat{\phi}}_{W_i}(W_i X) - \phi_{W_i,n}|_\infty \leq \sup_{\Omega_n} c |\phi'''_{W_i,n}|_\infty \delta_n^2$, so

$$\sup_{\Omega_n} \left| \int \overline{\hat{\phi}}_{W_i}(W_i X) X_k \, dP_n(X) - \int \phi_{W_i,n}(W_i X) X_k \, dP_n \right|$$
$$\tag{6.7}$$
$$\leq \sup_{\Omega_n} |\phi'''_{W_i,n}|_\infty \delta_n^2 \int |X_k| P_n$$
$$= o_P(1) \qquad \text{(by C6)}.$$

By C4, ULLN holds for $\{\phi_{W_i}(W_iX)X_k : W \in \Omega_n\}$, and by Lemma A.1, $\sup_{\Omega_n} P(W_iX \notin [\underline{b}_{ni}, \overline{b}_{ni}]) = o(1)$, so

$$(6.8) \qquad \sup_{\Omega_n} \left| \int \phi_{W_i}(W_iX)X_kI(W_iX \notin [\underline{b}_{ni}, \overline{b}_{ni}]) \, dP_n \right| = o_P(1).$$

Recall that $\phi_{W_i,n}(x) = \phi_{W_i}(x)I(x \in [\underline{b}_{ni}, \overline{b}_{ni}])$. From (6.6)–(6.8), we obtain

$$(6.9) \quad \sup_{\Omega_n} \left| \int \hat{\phi}_{W_i}(W_iX)X_k \, dP_n(X) - \int \phi_{W_i}(W_iX)X_k \, dP_n(X) \right| = o_P(1).$$

Now, by C4,

$$(6.10) \qquad \sup_{\Omega_n} \left| \int \phi_{W_i}(W_iX)X_k \, d(P_n - P) \right| = o_P(1),$$

and by continuity,

$$(6.11) \qquad \sup_{\Omega_n} \left| \int \phi_{W_i}(W_iX)X_k \, dP - \int \phi_{W_{P_i}}(W_PX)X_k \, dP \right| = o(1).$$

(6.3) then follows from (6.9)–(6.11).

In the following, we prove (6.4). Note that

$$(6.12) \quad \frac{\partial}{\partial W_{ik}}\{\hat{\phi}_{W_i}(W_iX)\} = \frac{\partial}{\partial W_{ik}}\{\gamma_n^T(W_i)\}\mathbf{B}_n(W_iX) + \hat{\phi}'_{W_i}(W_iX)X_k.$$

It suffices to show that the following hold:

$$[4] \quad \sup_{\Omega_n} \left| \int \hat{\phi}'_{W_i}(W_iX)X_kW_jX \, dP_n(X) - \int \phi'_{P_i}(W_{P_i}X)X_kW_{Pj}X \, dP \right| = o_P(1);$$

$$[5] \quad \sup_{\Omega_n} \left| \int \frac{\partial}{\partial W_{ik}}\{\gamma_n^T(W_i)\}\mathbf{B}_n(W_iX)W_jX \, dP_n(X) \right| = o_P(1).$$

Similarly to (6.3), the uniform convergence of [4] can be verified using conditions C4, C6, C7 and Lemmas A.1, A.2 and A.6. Further, the left-hand side of [5] is bounded by

$$\sup_{\Omega_n} \left\{ \left\| \frac{\partial}{\partial W_{ik}}\gamma_n(W_i) \right\|_2 \cdot \left\| \int \mathbf{B}_n(W_iX)W_jX \, dP_n \right\|_2 \right\} = O_P(\delta_n^{-7/2}n_i^{1/2}\varepsilon_n\delta_n^{-1}n_i^{1/2})$$
$$= o_P(1),$$

where the first equality follows from Lemma A.2 and Lemma A.4 and the second follows from C7. Thus, [5] holds and, hence, (6.4) is proved. $\square$

PROPOSITION 6.4. *Under the conditions of Theorem* 3.1, *condition* [V] *holds, that is,*

$$\sup_{W \in \Omega_n} \left| \int \hat{\mathbf{l}}^*\hat{\mathbf{l}}^{*T}(X, W) \, dP_n - \int \mathbf{l}^*\mathbf{l}^{*T}(X, W_P, \Phi_{W_P}) \, dP \right| = o_P(1).$$

PROOF. By checking the elements of $\hat{\mathbf{l}}^* \hat{\mathbf{l}}^{*T}(\mathbf{x}, W)$, it suffices to show that for $1 \leq i, j, k, k' \leq m$ and $i \neq j$,

$$\sup_{\Omega_n} \left| \int \hat{\phi}_{W_i}(W_i X) \hat{\phi}_{W_j}(W_j X) X_k X_{k'} \, dP_n \right.$$

$$\left. - \int \phi_{Pi}(W_{Pi} X) \phi_{Pj}(W_{Pj} X) X_k X_{k'} \, dP \right| = o_P(1),$$

$$\sup_{\Omega_n} \left| \int \hat{\phi}_{W_i}(W_i X) X_k X_{k'} \, dP_n(X) - \int \phi_{Pi}(W_{Pi} X) X_k X_{k'} \, dP \right| = o_P(1)$$

and

$$\sup_{\Omega_n} \left| \int \hat{\phi}_{W_i}(W_i X) X_k \kappa(W_j X) \, dP_n(X) - \int \phi_{Pi}(W_{Pi} X) X_k \kappa(W_{Pj} X) \, dP \right| = o_P(1).$$

Each of these can be verified using Lemmas A.1, A.2, A.6 and conditions C4, C6 and C7 with arguments similar to those used in proving (6.3). □

## 7. Conclusion.

In this paper, we viewed the classical ICA model within the framework of semiparametric models and obtained an asymptotically efficient estimator for the unmixing matrix by solving an approximate efficient score equation. The main difference between this new method and popular parametric ICA methods is that we estimate the density score functions of hidden sources adaptively. A variety of simulations have illustrated statistical efficiency of this estimator in comparison with state-of-the-art ICA algorithms.

## APPENDIX

**A.1. Some useful formulas.** Let $v = w W_P^{-1}$. Then $wX = vS$. If $v_k \neq 0$ for some $k \in \{1, \ldots, m\}$, then by the classical convolution formula, we have

$$f_w(t) = \int \frac{1}{v_k} r_k \left( \frac{t - \sum_{j \neq k} v_j s_j}{v_k} \right) \prod_{j \neq k} r_j(s_j) \, ds_j = E \left\{ \frac{1}{v_k} r_k \left( \frac{t - \sum_{j \neq k} v_j S_j}{v_k} \right) \right\}.$$

Since $f_w(t)$ is a marginal density function of $(vS, S_j : 1 \leq j \neq k \leq m)$, by a standard formula (see, e.g., [4]),

$$(A.1) \quad \phi_w(t) = -\frac{1}{v_k} E \left\{ \frac{r'_k}{r_k} \left( \frac{t - \sum_{j \neq k} v_j S_j}{v_k} \right) \Big| vS = t \right\} = \frac{1}{v_k} E[\phi_k(S_k) | vS = t]$$

and further calculation gives

$$(A.2) \quad \frac{\partial}{\partial t} \phi_w(t) = \phi_w^2(t) - \frac{1}{v_k^2} E \left\{ \frac{r''_k}{r_k} \left( \frac{t - \sum_{j \neq k} v_j S_j}{v_k} \right) \Big| vS = t \right\}.$$

**A.2. Calculation of the efficient score.** To formulate the tangent space defined in Section 2.1 for each nuisance parameter $r_i$, by taking the smooth submodel $\{r_i(\cdot; t) = r_i(\cdot) e^{t h_i(\cdot)} : |t| < 1\}$ for some $h_i \in \mathcal{L}^2(P)$, we have

$$\lim_{t \to 0} \frac{\partial}{\partial t} \{\log p_X(\mathbf{x}, W, r_1, \dots, r_i(\cdot, t), \dots, r_m)\} = h_i(W_i \mathbf{x}).$$

Since $r_i(\cdot; t)$ needs to be a probability density function which satisfies the mean and absolute median assumptions, $h_i$ needs to satisfy $E[h_i(S_i)] = 0$, $E[h_i(S_i)S_i] = 0$ and $E[h_i(S_i)\kappa(S_i)] = 0$, but is otherwise arbitrary. Thus, the tangent space for $r_i$ can be expressed as

$$TS_i = \{h_i(W_i\mathbf{x}) \in \mathcal{L}^2(P) | E[h_i(S_i)] = 0, E[h_i(S_i)S_i] = 0, E[h_i(S_i)\kappa(S_i)] = 0\}.$$

Note that the tangent spaces $\{TS_i : 1 \le i \le m\}$ are perpendicular to each other since the $S_i$ are mutually independent. Thus, any projection onto the tangent space of $(r_1, \dots, r_m)$ is equal to the summation of the partial projection onto each $TS_i$. The efficient score of $W$ becomes

$$\mathbf{l}^*(., W, \Phi) = \dot{\mathbf{l}}_W - \sum_{i=1}^m \pi(\dot{\mathbf{l}}_W \mid TS_i),$$

where $\pi(.|L)$ denotes the projection operator in $\mathcal{L}^2(P_{(W, r_1, \dots, r_m)})$ onto $L$. Since each off-diagonal entry of $I_{m \times m} - \Phi(S)S^T$ is perpendicular to all $TS_i$ and each diagonal entry of it is perpendicular to all but one $TS_i$, $\mathbf{l}^*$ can be obtained as in (2.4)–(2.8) of Section 2.3 by using the fact that $TS_i^\perp = span\{1, S_i, \kappa(S_i)\}$.

**A.3. Some properties of cubic B-splines.** Let $\xi_1 < \xi_2 < \cdots < \xi_N$ be fixed points. The first order B-spline basis functions based on these knots can be expressed as $B_i^1(x) = I(x \in [\xi_i, \xi_{i+1}))$, $i = 1, \dots, N-1$, and the $k$th order B-spline basis functions can be obtained recursively $(k \ge 2)$ by

$$B_i^k(x) = \frac{x - \xi_i}{\xi_{i+k-1} - \xi_i} B_i^{k-1}(x) + \frac{\xi_{i+k} - x}{\xi_{i+k} - \xi_{i+1}} B_{i+1}^{k-1}(x),$$

for $i = 1, \dots, N-k$. Each $B_i^k(x)$ is differentiable w.r.t. $x$ up to order $k - 2$ and its first order derivative can be expressed as

$$\frac{d}{dx} B_i^k(x) = \frac{k-1}{\xi_{i+k-1} - \xi_i} B_i^{k-1}(x) - \frac{k-1}{\xi_{i+k} - \xi_{i+1}} B_{i+1}^{k-1}(x).$$

We use the 4th order, so-called cubic, B-splines $\{B_i^4 : 1 \le i \le N-4\}$ with equally spaced knots, that is, $\xi_{i+1} - \xi_i = \delta$ $(i = 1, \dots, N-1)$ for some algorithm-determined $\delta$. For simplicity, the superscript in $B_i^4$ is omitted below. The following properties of cubic B-splines will be frequently used in proving the lemmas below (see [11, 23] for details):

(I)  $0 \le B_i(x) < 1$, $B_i(x)B_j(x) = 0$ if $|i - j| > 3$.

(II)  $|\frac{d}{dx}B_i(x)| < \delta^{-1}$, $|\frac{d^2}{dx^2}B_i(x)| < 2\delta^{-2}$.

(III)  $\sum_{i=1}^{N}[B_i(x)]^2 < 1$, $\sum_{i=1}^{N}[\frac{d^2}{dx^2}B_i(x)]^2 < 6\delta^{-4}$.

**A.4. Supporting lemmas for Propositions 6.1–6.4.**  In this subsection, we prove all of the lemmas used in the proofs of Propositions 6.1–6.4. Recall that for each $\phi_k$ $(k = 1, \ldots, m)$, we have an interval $[\underline{b}_{nk}, \overline{b}_{nk}]$ and $n_k$ cubic B-spline basis functions defined thereupon using equally spaced knots on it, say $\mathbf{B}_n^{(k)} = (B_{n1}^{(k)}, \ldots, B_{nn_k}^{(k)})^T$, as in Section 2.4. Thus, we have constructed a sequence of sieves $\mathcal{G}_n^{(k)}$ using $\mathbf{B}_n^{(k)}$ as basis functions. For any $W \in \Omega_n$, we have a class of estimates $\hat{\phi}_{W_k} \in \mathcal{G}_n^{(k)}$ for $\phi_{W_k}$, as defined in Section 2.4. For convenience, the superscript of $\mathbf{B}_n^{(k)}$ is often omitted below.

Let $\Omega_n^{(k)} = \{W_k : W \in \Omega_n\}$ for $k = 1, \ldots, m$. We also need an intermediate approximation function $\overline{\overline{\phi}}_{W_k} \in \mathcal{G}_n^{(k)}$, defined as follows. For $w \in \Omega_n^{(k)}$,

$$(A.3) \qquad \overline{\overline{\phi}}_w = \gamma(w)^T \mathbf{B}_n^{(k)},$$

where $\gamma(w) = A(w)^{-1}D(w)$ with $A(w) = \int \mathbf{B}_n^{(k)} \mathbf{B}_n^{(k)T}(wX)\,dP$ and $D(w) = \int [\mathbf{B}_n^{(k)}]'(wX)\,dP$. Note that the subscript $w$ of $\overline{\overline{\phi}}_w$ should always be associated with $\Omega_n^{(k)}$ for some $k \in \{1, \ldots, m\}$, similarly for $\hat{\phi}_w$.

In the following, $c$ denotes a constant (only dependent on the population law $P$), but its exact value may vary in different places (even in a single line) without clarification. For a column vector $\mathbf{x} \in \mathbb{R}^m$, $\|\mathbf{x}\|_2 = \sqrt{\mathbf{x}^T\mathbf{x}}$. For an $m \times m$ real matrix $A$, $\|A\|_1 = \max_{1 \le i \le m} \|A_i\|_2$, $\|A\|_2 = \max_{x \in R^m, \|x\|_2 = 1} \|Ax\|_2$ and $\|A\|_F = \sqrt{tr(A^TA)}$, thus, $\|A\|_2 \le \|A\|_1$.

The following Lemmas A.1–A.8 hold under the conditions of Theorem 3.1. Jin [17] obtained results similar to Lemma A.2 and Lemmas A.5–A.7 concerning the B-spline approximation, but in generally different settings.

LEMMA A.1.  *For sufficient large $n$*, $\sup_{w \in \Omega_n^{(k)}} |f_w|_\infty < \infty$, $\sup_{w \in \Omega_n^{(k)}} |f_w'|_\infty < \infty$, $\sup_{w \in \Omega_n^{(k)}} \min_{t \in [\underline{b}_{nk}, \overline{b}_{nk}]} f_w(t) \ge c\delta_n$ *and* $\sup_{w \in \Omega_n^{(k)}} P(wX \notin [\underline{b}_{ni}, \overline{b}_{ni}]) = o(1)$.

PROOF.  The first two inequalities follow easily from C3. The remaining two are proved as follows. For any $w \in \Omega_n^{(k)}$, $\|w - W_{Pk}\|_2 \le \varepsilon_n$. If we let $v = wW_P^{-1}$, then $|v_j| \to 0$ for $1 \le j \ne k \le m$ and $v_k \to 1$ as $n \to \infty$. Since $f_w(t) = E[\frac{1}{v_k}r_k(\frac{t - \sum_{j \ne k} v_j S_j}{v_k})]$, we can consider the right-hand side as a function (say

$h$) of $v$. By the first order Taylor expansion,

$$|f_w(t) - r_k(t)| \leq \varepsilon_n \|W_P^{-1}\|_2 \left\{ \sum_{j=1}^m \sup_{w \in \Omega_n^{(k)}} \left| \frac{\partial}{\partial v_j} h(v) \right| \right\} \leq c\varepsilon_n,$$

where by direct calculation and using C3, $|\frac{\partial}{\partial v_j} h(v)|$ is uniformly bounded with $w \in \Omega_n^{(k)}$ and $t \in \mathbb{R}$. Recall that $r_k(t) \geq c\delta_n$ for $t \in [\underline{b}_{nk}, \overline{b}_{nk}]$ and $\varepsilon_n \ll \delta_n$, thus

$$\sup_{w \in \Omega_n^{(k)}} \min_{t \in [\underline{b}_{nk}, \overline{b}_{nk}]} f_w(t) \geq c\delta_n.$$

Finally,

$$P(wX \in [\underline{b}_{ni}, \overline{b}_{ni}]) = \int_{[\underline{b}_{nk}, \overline{b}_{nk}]} f_w(t)\,dt \geq \int_{[\underline{b}_{nk}, \overline{b}_{nk}]} (r_k(t) - c\varepsilon_n)\,dt$$

$$= P(S_k \in [\underline{b}_{nk}, \overline{b}_{nk}]) - c\varepsilon_n(\overline{b}_{nk} - \underline{b}_{nk}).$$

Since $\varepsilon_n(\overline{b}_{nk} - \underline{b}_{nk}) = o(1)$ and $P(S_k \in [\underline{b}_{nk}, \overline{b}_{nk}]) \to 1$, we have $\inf_{w \in \Omega_n^{(k)}} P(wX \in [\underline{b}_{ni}, \overline{b}_{ni}]) \to 1$.  □

Recall the definitions of $\hat{\phi}_{W_k}$, $\gamma_n(W_k)$, $A_n(W_k)$ and $D_n(W_k)$ in Section 2.4 and that of $\gamma(w) = [A(w)]^{-1}D(w)$ in (A.3). The lemmas below give their uniform convergence rates.

LEMMA A.2.  *The following holds for $k, i = 1, \ldots, m$:*

(1) $\|D(w)\|_2 \leq c\delta_n n_k^{1/2}$; $c\delta_n^2 \leq eig(A(w)) \leq c\delta_n$ for $w \in \Omega_n^{(k)}$;

(2) $\sup_{w \in \Omega_n^{(k)}} \|D_n(w) - D(w)\|_2 = O_P((n_k \log n_k)^{1/2}(n\delta_n)^{-1/2})$;

(3) $\sup_{w \in \Omega_n^{(k)}} \|D_n(w)\|_2 = O_P(\delta_n n_k^{1/2})$;

(4) $\sup_{w \in \Omega_n^{(k)}} \|A_n(w) - A(w)\|_2 = O_P((\delta_n \log n_k)^{1/2}n^{-1/2})$;

(5) $\sup_{w \in \Omega_n^{(k)}} \|A_n^{-1}(w)\|_2 = O_P(\delta_n^{-2})$;

(6) $\sup_{w \in \Omega_n^{(k)}} \|\gamma_n(w)\|_2 = O_p(n_k^{1/2}\delta_n^{-1})$;

(7) $\sup_{\Omega_n^{(k)}} \|\gamma_n(w) - \gamma(w)\|_2 = o_P(1)$;

(8) $\sup_{\Omega_n^{(k)}} \|\frac{\partial}{\partial w_i}\{A_n(w)\}\|_2 = O_P(\delta_n^{-1/2})$;

(9) $\sup_{\Omega_n^{(k)}} \|\frac{\partial}{\partial w_i}\{D_n(w)\}\|_2 = O_P(\delta_n^{-2})$;

(10) $\sup_{\Omega_n^{(k)}} \|\frac{\partial}{\partial w_i}\{\gamma_n(w)\}\|_2 = o_P(\delta_n^{-7/2}n_k^{1/2})$.

PROOF.  The procedure is as follows. First, (3) is implied by (1) and (2). Second, (6) is implied by (3) and (5). Third, since

$$\frac{\partial}{\partial w_i}\{\gamma_n(w)\} = -A_n^{-1}(w)\frac{\partial}{\partial w_i}\{A_n(w)\}\gamma_n(w) + A_n^{-1}(w)\frac{\partial}{\partial w_i}\{D_n(w)\},$$

(10) is implied by (5), (6), (8) and (9). Hence, it suffices to prove (1), (2), (4), (5), (7), (8) and (9). Further, the proofs of (2) and (4) are similar and the proofs of (8) and (9) are similar. Thus the proofs of (4) and (9) are omitted.

PROOF OF (1). By taking the derivatives of the cubic B-splines, $B'_{ni}(t) = \delta_n^{-1}(B_{ni}^3(t) - B_{n,i+1}^3(t))$, where $B_{ni}^3$ are the third order B-splines defined on the same knots, $i = 1, \ldots, n_k$, we have

$$
\begin{aligned}
|D_i(w)| &= \delta_n^{-1} \left| \int (B_{n,i}^3(t) - B_{n,i+1}^3(t)) f_w(t) \, dt \right| \\
&= \delta_n^{-1} \left| \int B_{n,i}^3(t)(f_w(t) - f_w(t + \delta_n)) \, dt \right| \\
&\leq \int B_{n,i}^3(t) \, dt |f'_w|_\infty < 3\delta_n |f'_w|_\infty.
\end{aligned}
$$

So the first result holds due to Lemma A.1. The second result follows from Lemma 5.1 of [17]. □

PROOF OF (2). Note that

$$
\begin{aligned}
&P\Big( \sup_{w \in \Omega_n^{(k)}} \|D_n(w) - D(w)\|_2 \geq t \Big) \\
&\quad = P\Big( \sup_{w \in \Omega_n^{(k)}} \Big\| \int \mathbf{B}'_n(wX) \, d(P_n - P) \Big\|_2 \geq t \Big) \\
&\quad \leq \sum_{i=1}^{n_k} P\Big( \sup_{w \in \Omega_n^{(k)}} \Big| \int B'_{n,i}(wX) \, d(P_n - P) \Big| \geq \frac{t}{\sqrt{n_k}} \Big).
\end{aligned}
$$

For a fixed pair $(i, k)$, let $\mathcal{F}_n = \{g_w(\mathbf{x}) = B'_{n,i}(w\mathbf{x}) : w \in \Omega_n^{(k)}\}$, a class of functions indexed by $\Omega_n^{(k)}$. Then for $w, v \in \Omega_n^{(k)}$, $\|g_w(\mathbf{x}) - g_v(\mathbf{x})\|_2 \leq 2\delta_n^{-2}\|w - v\|_2 \|\mathbf{x}\|_2$ by property II of cubic B-splines. Now, the index set $\Omega_n^{(k)}$ can be covered by $N = c\varepsilon_n^m u^{-m}$ balls of radius $u$ and for any $w, v$ in the same ball, $E[\|g_w(X) - g_v(X)\|_2] \leq c\delta_n^{-2}u$. Further, by property I of cubic B-splines, $\sup_{g_w \in \mathcal{F}_n} |g_w|_\infty \leq \delta_{nk}^{-1}$. Then for $c \max(\delta_n^{-1}, \varepsilon_n \delta_n^{-2}) \leq a \leq c\sqrt{n}$,

$$
P\Big( \sup_{w \in \Omega_n^{(k)}} \sqrt{n} \Big| \int B'_{n,i}(wX) \, d(P_n - P) \Big| \geq a \Big) \leq \exp(-ca^2 \delta_n).
$$

This follows from Theorem 5.11 of van de Geer ([24], page 75) which generalizes Hoeffding's inequality and calculates the uniform convergence rate for a class of functions in terms of its size measure, or so-called *bracketing entropy*; see [24] for details. Note that $\varepsilon_n \ll \delta_n$ by C7, so for $t \geq c n_k^{1/2} n^{-1/2} \delta_n^{-1}$, we have

$$P\Big(\sup_{w \in \Omega_n^{(k)}} \|D_n(w) - D(w)\|_2 \geq t\Big) \leq n_k \exp(-ct^2 n \delta_n n_k^{-1}).$$

Thus,

$$\sup_{\Omega_n^{(k)}} \|D_n(w) - D(w)\|_2 = O_p\bigg(\sqrt{\frac{n_k \log n_k}{n \delta_n}}\bigg). \qquad \square$$

PROOF OF (5). Since $A_n^{-1} = (A + A_n - A)^{-1} = A^{-1}(I + (A_n - A)A^{-1})^{-1}$, by (1) and (4) (omitting the index $w$), we have

$$\sup_{w \in \Omega_n^{(k)}} \|A_n - A\|_2 \|A^{-1}\|_2 = o_p(1),$$

hence,

$$\sup_{w \in \Omega_n^{(k)}} \|A_n^{-1}\|_2 \leq \sup_{w \in \Omega_n^{(k)}} \|A^{-1}\|_2 (1 - \|A_n - A\|_2 \|A^{-1}\|_2)^{-1} = O_P(\delta_n^{-2}).$$

Here, we use the inequality of matrix norm $\|(I + A)^{-1}\|_2 \leq (1 - \|A\|_2)^{-1}$ for any square matrix $A$ with $\|A\|_2 < 1$, where $I$ is the identity matrix. $\square$

PROOF OF (7). Since by (1)–(5),

$$\sup_{w \in \Omega_n^{(k)}} \|\gamma_n(w) - \gamma(w)\|_2 = \sup_{w \in \Omega_n^{(k)}} \|A^{-1}(D_n - D) - A_n^{-1}(A_n - A)A^{-1}D_n\|_2$$

$$= O_p\bigg(\delta_n^{-2}\sqrt{\frac{n_k \log n_k}{n \delta_n}} + \delta_n^{-2}\sqrt{\frac{\delta_n \log n_k}{n}}\delta_n\sqrt{n_k}\delta_n^{-2}\bigg)$$

$$= o_P(1),$$

where the last equality follows from C7. $\square$

PROOF OF (8). Note that the partial derivative is (omitting ($i$) from $\mathbf{B}_n^{(i)}$),

$$\frac{\partial}{\partial w_k}\{A_n(w)\}_{jj'} = \int (B_{nj}B_{nj'}' + B_{nj}'B_{nj'})(wX)X_k\,dP_n.$$

By the Cauchy–Schwarz inequality,

$$\bigg|\int (B_{nj}B_{nj'}' + B_{nj}'B_{nj'})(wX)X_k\,dP_n\bigg|^2$$

$$\leq \int (B_{nj}B_{nj'}' + B_{nj}'B_{nj'})^2(wX)\,dP_n \int X_k^2\,dP_n.$$

Similarly, we can calculate the uniform convergence rate,

$$\sup_{0 \le j,j' \le n_i, |j-j'| \le 3} \sup_{w \in \Omega_n^{(i)}} \int (B_{nj}B'_{nj'} + B'_{nj}B_{nj'})^2 (wX)\, d(P_n - P) = o_p(1).$$

Further, from Lemma A.1, $\sup_{w \in \Omega_n^{(k)}} |f_w|$ is bounded so after algebraic expansion, we have

$$\sup_{w \in \Omega_n^{(i)}} \int (B_{nj}B'_{nj'} + B'_{nj}B_{nj'})^2 (wX)\, dP \le c\delta_n^{-1}.$$

Thus, $|\frac{\partial}{\partial w_k}\{A_n(w)\}_{jj'}| \le c\delta_n^{-1/2}$. For cubic B-splines, $B_{nj}(x)B'_{nj'}(x) \equiv 0$ for $|j - j'| > 3$, thus, each row of $\frac{\partial}{\partial w_k}\{A_n(w)\}$ has at most seven nonzero elements. So,

$$\sup_{\Omega_n^{(i)}} \left\|\frac{\partial}{\partial w_k}\{A_n(w)\}\right\|_2 \le \sup_{\Omega_n^{(i)}} \left\|\frac{\partial}{\partial w_k}\{A_n(w)\}\right\|_1 = O_p(\delta_n^{-\frac{1}{2}}). \qquad \square$$

LEMMA A.3. $\|\int \mathbf{B}_n^{(i)}(S_i)S_j\, dP_n\|_2 = O_P(n^{-1/2})$, where $S_i = W_{Pi}X$, $1 \le i \ne j \le m$.

PROOF. (We omit $(i)$ in $\mathbf{B}_n^{(i)}, B_{nk}^{(i)}$.)

$$E\left(\left\|\int \mathbf{B}_n(S_i)S_j\, dP_n\right\|_2^2\right) = \frac{1}{n}E\left(\sum_{k=1}^{n_i} B_{nk}(S_i)^2 S_j^2\right) \le \frac{4}{n}E(S_j^2). \qquad \square$$

LEMMA A.4. $\sup_{\Omega_n} \|\int \mathbf{B}_n^{(i)}(W_iX)W_jX\, dP_n\|_2 = O_p(\varepsilon_n\delta_n^{-1}n_i^{-1/2})$ for $1 \le i \ne j \le m$.

PROOF. Similarly to the proof of Lemma A.2(2), we have

$$\sup_{\Omega_n} \left\|\int \mathbf{B}_n(W_iX)W_jX\, d(P_n - P)\right\|_2 = O_P\left(\sqrt{\frac{\delta_n^2 n_i \log n_i}{n}}\right).$$

Further, note that $|B_{nk}(x) - B_{nk}(y)| \le \delta_n^{-1}|x - y|$, so for any $W \in \Omega_n$,

$$\left\|\int \mathbf{B}_n(W_iX)W_jX\, dP\right\|_2^2$$
$$= \sum_{k=1}^{n_i} \left|\int (B_{nk}(W_iX)W_jX - B_{nk}(W_{Pi}X)W_{Pj}X)\, dP\right|^2$$
$$\le \sum_{k=1}^{n_i} (\delta_n^{-1}E\|X\|_2^2\|W_i - W_{Pi}\|_2\|W_i\|_2 + E\|X\|_2\|W_i - W_{Pi}\|_2)^2$$
$$\le c\varepsilon_n^2\delta_n^{-2}n_i.$$

Thus, $\sup_{\Omega_n} \| \int \mathbf{B}_n^{(i)}(W_i X) W_j X \, dP_n \|_2 = O_P(\sqrt{\frac{\delta_n^2 n_i \log n_i}{n}} + \varepsilon_n \delta_n^{-1} n_i^{1/2})$.   $\square$

LEMMA A.5.   $E\{(\overline{\tilde{\phi}}_{W_i}(W_i X) - \phi_{W_i,n}(W_i X))^2\} \leq \delta_n^6 |\phi_{W_i,n}'''|_\infty^2$.

PROOF.   Let $d(\phi_{W_i,n}, \mathcal{G}_n) = \inf_{h \in \mathcal{G}_n} |\phi_{W_i,n} - h|_\infty$. Then by the Jackson-type theorem [11],

$$d(\phi_{W_i,n}, \mathcal{G}_n) \leq c\delta_n^3 |\phi_{W_i,n}'''|_\infty.$$

Thus, the result follows from

$$\begin{aligned}
E\{(\overline{\tilde{\phi}}_{W_i}(W_i X) - \phi_{W_i,n}(W_i X))^2\} &= \inf_{h \in \mathcal{G}_n^{(i)}} E\{(h(W_i X) - \phi_{W_i,n}(W_i X))^2\} \\
&\leq d(\phi_{W_i,n}, \mathcal{G}_n)^2. \qquad\qquad \square
\end{aligned}$$

LEMMA A.6.   $|\overline{\tilde{\phi}}_{W_i} - \phi_{W_i,n}|_\infty \leq c\delta_n^2 |\phi_{W_i,n}'''|_\infty$; $|\overline{\tilde{\phi}}_{W_i}' - \phi_{W_i,n}'|_\infty \leq c|\phi_{W_i,n}'''|_\infty \delta_n$.

PROOF.   By Theorem XII.4 of de Boor (1978), there exists a quasi-interpolant with some $a \in \mathbb{R}^{n_i}$,

$$\tilde{\tilde{\phi}}_{W_i}(t) = a^T \mathbf{B}_n(t),$$

such that $\tilde{\tilde{\phi}}_{W_i}$ simultaneously approximates $\phi_{W_i,n}$ and its first derivative to optimal order, that is,

$$|\tilde{\tilde{\phi}}_{W_i} - \phi_{W_i,n}|_\infty = c|\phi_{W_i,n}'''|_\infty \delta_n^3 \quad \text{and} \quad |\tilde{\tilde{\phi}}_{W_i}' - \phi_{W_i,n}'|_\infty = c|\phi_{W_i,n}'''|_\infty \delta_n^2.$$

So,

$$E(\tilde{\tilde{\phi}}_{W_i}(W_i X) - \phi_{W_i,n}(W_i X))^2 \leq c|\phi_{W_i,n}'''|_\infty^2 \delta_n^6.$$

Combining this with Lemma A.5, we have

$$E(\overline{\tilde{\phi}}_{W_i} - \tilde{\tilde{\phi}}_{W_i})^2 \leq E(\tilde{\tilde{\phi}}_{W_i} - \phi_{W_i,n})^2 + E(\overline{\tilde{\phi}}_{W_i} - \phi_{W_i,n})^2 \leq c|\phi_{W_i,n}'''|_\infty^2 \delta_n^6.$$

Let $coef(\tilde{\tilde{\phi}}_{W_i})$ and $coef(\overline{\tilde{\phi}}_{W_i})$ be coefficients of $\mathbf{B}_n$ in $\tilde{\tilde{\phi}}_{W_i}$ and $\overline{\tilde{\phi}}_{W_i}$, respectively. Then

$$\begin{aligned}
E(\overline{\tilde{\phi}}_{W_i} - \tilde{\tilde{\phi}}_{W_i})^2 &= E((coef(\tilde{\tilde{\phi}}_{W_i}) - coef(\overline{\tilde{\phi}}_{W_i}))^T \mathbf{B}_n)^2 \\
&\geq \lambda_n \|coef(\tilde{\tilde{\phi}}_{W_i}) - coef(\overline{\tilde{\phi}}_{W_i})\|_2^2,
\end{aligned}$$

where $\lambda_n$ is the minimum eigenvalue of $A(W_i) = E[\mathbf{B}_n \mathbf{B}_n^T(W_i X)]$. By Lemma A.2, $\lambda_n \geq c\delta_n^2$. Thus,

$$|\overline{\tilde{\phi}}_{W_i} - \tilde{\tilde{\phi}}_{W_i}|_\infty \leq \|coef(\tilde{\tilde{\phi}}_{W_i}) - coef(\overline{\tilde{\phi}}_{W_i})\|_2 \leq c|\phi_{W_i,n}'''|_\infty \delta_n^2.$$

Hence,

$$\sup_{\Omega_n} |\overline{\hat{\phi}}_{W_i} - \phi_{W_i,n}| \le \sup_{\Omega_n} c |\phi'''_{W_i,n}|_\infty \delta_n^2.$$

Further, by observing that $|B'_{nk}|_\infty \le \delta_n^{-1}$, we have

$$|\overline{\hat{\phi}}'_{W_i} - \tilde{\hat{\phi}}'_{W_i}|_\infty \le \|coef(\tilde{\hat{\phi}}_{W_i}) - coef(\overline{\hat{\phi}}_{W_i})\|_2 \delta_n^{-1} \le c |\phi'''_{W_i,n}|_\infty \delta_n.$$

Thus,

$$|\overline{\hat{\phi}}'_{W_i} - \phi'_{W_i,n}|_\infty \le |\tilde{\hat{\phi}}'_{W_i} - \phi'_{W_i,n}|_\infty + |\overline{\hat{\phi}}'_{W_i} - \tilde{\hat{\phi}}'_{W_i}|_\infty \le c |\phi'''_{W_i,n}|_\infty \delta_n. \qquad \square$$

LEMMA A.7. $\int (\hat{\phi}_{W_{Pk}}(S_k) - \phi_k(S_k))^2 \, dP_n = o_p(1).$

PROOF. Observe that

$$\frac{1}{3} \int (\hat{\phi}_{W_{Pk}}(S_k) - \phi_k(S_k))^2 \, dP_n \le \int (\hat{\phi}_{W_{Pk}}(S_k) - \overline{\hat{\phi}}_{W_{Pk}}(S_k))^2 \, dP_n$$
$$+ \int (\overline{\hat{\phi}}_{W_{Pk}}(S_k) - \phi_{k,n}(S_k))^2 \, dP_n$$
$$+ \int \phi_k(S_k)^2 I(S_k \notin [\underline{b}_{nk}, \overline{b}_{nk}]) \, dP_n.$$

The remainder of the proof is similar to the use of (6.1) in proving Proposition 6.1 by using Lemma A.2 and Lemma A.6. $\square$

LEMMA A.8. *Let* $\{p(\cdot, \theta, \eta) : \theta \in \Omega \subset \mathbb{R}^d, \eta \in \mathcal{E}\}$ *be a parametric or semiparametric model, where* $\theta$ *is the parameter of interest. Suppose that moderate regularity conditions are satisfied and that* $\mathbf{l}^*(\cdot, \theta, \eta)$ *is the efficient score function of* $\theta$*, as defined in* [5]*. Then*

$$\int \frac{\partial}{\partial \theta} \mathbf{l}^*(X, \theta, \eta) \, dP_{(\theta,\eta)} = - \int \mathbf{l}^* \mathbf{l}^{*T}(X, \theta, \eta) \, dP_{(\theta,\eta)}.$$

PROOF. We only prove this for the parametric case $\mathcal{E} \subset \mathbb{R}^m$. Let $\mathbf{I}(\theta, \eta)$ be the information matrix of $(\theta, \eta)$. Then by classic likelihood theory (e.g., Proposition 2.4.1 of [5]), $\mathbf{l}^*(\cdot, \theta, \eta) = \dot{\mathbf{l}}_1 - (\mathbf{I}_{12} \mathbf{I}_{22}^{-1})(\theta, \eta) \dot{\mathbf{l}}_2$. Here, $\dot{\mathbf{l}}_1$ and $\dot{\mathbf{l}}_2$ are the partial derivatives of $l(\cdot, \theta, \eta) \equiv \log p(\cdot, \theta, \eta)$ w.r.t. $\theta$ and $\eta$, respectively. Similarly, $\ddot{\mathbf{l}}_{ij}$ $(i, j = 1, 2)$ are defined as second order derivatives of $l(\cdot, \theta, \eta)$ w.r.t. $(\theta, \eta)$. Thus, $\frac{\partial}{\partial \theta} \mathbf{l}^*(X, \theta, \eta) = \ddot{\mathbf{l}}_{11} - (\mathbf{I}_{12} \mathbf{I}_{22}^{-1})(\theta, \eta) \ddot{\mathbf{l}}_{21} - \frac{\partial}{\partial \theta} \{(\mathbf{I}_{12} \mathbf{I}_{22}^{-1})(\theta, \eta)\} \dot{\mathbf{l}}_2$. Since $\int \dot{\mathbf{l}}_2(X, \theta, \eta) dP_{(\theta,\eta)} = 0$, we have

$$\int \frac{\partial}{\partial \theta} \mathbf{l}^*(X, \theta, \eta) \, dP_{(\theta,\eta)} = \int \ddot{\mathbf{l}}_{11} \, dP_{(\theta,\eta)} - (\mathbf{I}_{12} \mathbf{I}_{22}^{-1})(\theta, \eta) \int \ddot{\mathbf{l}}_{21} \, dP_{(\theta,\eta)}.$$

Since the information matrix satisfies $\mathbf{I}_{ij} = -\int \ddot{\mathbf{l}}_{ij}(X, \theta, \eta) \, dP_{(\theta, \eta)}$, $i, j = 1, 2$, the result follows by $\int \mathbf{l}^* \mathbf{l}^{*T} dP = \mathbf{I}_{11} - \mathbf{I}_{12} \mathbf{I}_{22}^{-1} \mathbf{I}_{21}$ (see Proposition 2.4.1 of [5], page 32). For the semiparametric case, the reader is referred to [5] for a generalization of this proof.  $\square$

**Acknowledgments.** This paper was presented at the Joint Statistical Meeting in San Francisco, USA, August 2003. The results were reported in Chen's Ph.D. Thesis. We would like to thank Francis Bach at UC-Berkeley for sharing his simulation code and Scott Vander Wiel for helpful comments. We also wish to thank the Editors and four referees for useful suggestions that have helped significantly to improve the paper.

600 MOUNTAIN AVENUE, ROOM 2C-281
MURRAY HILL, NEW JERSEY 07974
USA
E-MAIL: aychen@research.bell-labs.com

367 EVANS HALL
BERKELEY, CALIFORNIA 94720
USA
E-MAIL: bickel@stat.berkeley.edu